\documentclass[journal]{IEEEtran}
\usepackage{blindtext}
\usepackage{cite}
\usepackage[pdftex]{graphicx}
\DeclareGraphicsExtensions{.pdf,.jpeg,.png}
\usepackage{algorithm}
\usepackage[tight,footnotesize]{subfigure}
\usepackage{caption}
\usepackage{color}
\usepackage[font=footnotesize]{subfig}
\usepackage{url}
\usepackage{amssymb}
\usepackage{amsmath}
\usepackage{mathtools}
\usepackage{makecell}
\usepackage{xcolor}
\usepackage{tikz}
\usepackage{booktabs}
\usepackage{multirow}
\usepackage{algorithm,algpseudocode}
\usepackage{amsmath}
\usepackage{amsfonts}
\usepackage{amssymb}
\usepackage{colortbl}
\usepackage{makecell}

\newcolumntype{P}[1]{>{\centering\arraybackslash}p{#1}}
\newcolumntype{M}[1]{>{\centering\arraybackslash}m{#1}}
\newcolumntype{N}{@{}m{0pt}@{}}

\makeatletter
\def\BState{\State\hskip-\ALG@thistlm}
\makeatother

\ifCLASSINFOpdf
  \usepackage[pdftex]{graphicx}
\else
\fi
\usepackage{amsmath}
\usepackage{url}

\begin{document}
%
\title{FantastIC4: A Hardware-Software Co-Design Approach for Efficiently Running 4bit-Compact Multilayer Perceptrons}
%
%
%

\author{
\IEEEauthorblockN{Simon Wiedemann$^\dag$, Suhas Shivapakash$^\dag$,  ~\IEEEmembership{Student Member,~IEEE}, Pablo Wiedemann$^\dag$, Daniel Becking, Wojciech Samek,~\IEEEmembership{Member,~IEEE}, Friedel Gerfers, ~\IEEEmembership{Member,~IEEE}, and  Thomas Wiegand,~\IEEEmembership{Fellow,~IEEE} }
\thanks{$^\dag$Equal contribution. }
\thanks{Suhas Shivapakash and Friedel Gerfers are with chair of Mixed Signal Circuit Design, Department
of Computer Engineering and Microelectronics, Technical University of Berlin, Berlin, Germany, e-mail: suhas.shivaprakash@tu-berlin.de}
\thanks{Simon Wiedemann, Pablo Wiedemann, Daniel Becking, Wojciech Samek are with Machine Learning Group,  Fraunhofer  Heinrich  Hertz  Institute,  Berlin, Germany, e-mail: wojciech.samek@hhi.fraunhofer.de}
\thanks{Thomas Wiegand is with chair of Media Technology, Technical University of Berlin and Fraunhofer  Heinrich  Hertz  Institute,  Berlin, Germany, e-mail: thomas.wiegand@hhi.fraunhofer.de }
}


%
%

\markboth{}%
{Wiedemann and Shivapakash \MakeLowercase{\textit{et al.}}: FantastIC4: A Hardware-Software Co-Design Approach for Efficiently Running 4bit-Compact Multilayer Perceptrons}
%



\maketitle

\begin{abstract}
With the growing demand for deploying deep learning models to the ``edge'', it is paramount to develop techniques that allow to execute state-of-the-art models within very tight and limited resource constraints. In this work we propose a software-hardware optimization paradigm for obtaining a highly efficient execution engine of deep neural networks (DNNs) that are based on fully-connected layers. Our approach is centred around compression as a means for reducing the area as well as power requirements of, concretely, multilayer perceptrons (MLPs) with high predictive performances. Firstly, we design a novel hardware architecture named \textit{FantastIC4}, which (1) supports the efficient on-chip execution of multiple compact representations of fully-connected layers and (2) minimizes the required number of multipliers for inference down to only 4 (thus the name). Moreover, in order to make the models amenable for efficient execution on FantastIC4, we introduce a novel entropy-constrained training method that renders them to be robust to 4bit quantization and highly compressible in size simultaneously. The experimental results show that we can achieve throughputs of 2.45 TOPS with a total power consumption of 3.6W on a Virtual Ultrascale FPGA XCVU440 device implementation, and achieve a total power efficiency of 20.17 TOPS/W on a 22nm process ASIC version. When compared to the other state-of-the-art accelerators designed for the Google Speech Command (GSC) dataset, FantastIC4 is better by 51$\times$ in terms of throughput and 145$\times$ in terms of area efficiency (GOPS/W).

\end{abstract}


\IEEEpeerreviewmaketitle

\section{Introduction}
In recent years, the topic of ``edge'' computing has gained significant attention due to the benefits that come along with processing data directly at its source of collection \cite{DL_edge_survey}. For instance, by running machine learning algorithms directly at the edge-device (e.g., wearables), latency issues can be greatly mitigated and/or increased privacy can be guaranteed since no data must be send to third-party cloud providers. Naturally, this has triggered the interest in deploying deep learning models to such embedded devices due to their high predictive performance. However, traditional deep learning models are usually very resource hungry since they entail a large number of parameters. In particular, processing a high number of parameters usually requires expensive hardware components such as large memory units and, if high throughput and low latency is desired, a high number of multipliers for parallel processing. This comes at the expense of spending lots of resources in power consumption and chip-area, thus greatly limiting their application in use-cases with tight area and power consumption budgets such as in the IoT or wearables.

This motivates the research of methods that can highly compress the DNNs weight parameters since, by doing so, we do not only minimize the respective data movement and therefore their power consumption, but also the required chip-area during execution. However, the efficient processing of compressed representations of data comes with a series of challenges, inter alia bit-alignment problems, reduction of locality, increased serialization, etc. Moreover, state-of-the-art compression techniques require complex decoding prior to performing arithmetic operations, which can compensate for the savings attained from compression specially when the hardware is not tailored to such type of decoding algorithms. This motivates a hardware-software co-design paradigm where, on the one hand, novel training techniques that make DNNs highly compressible are proposed and, on the other hand, novel hardware architectures are designed supporting the efficient, on-chip execution of compressed representations.

In this work we propose a software-hardware optimization paradigm which allows to efficiently execute highly compact representations of DNNs based on fully-connected (FC) layers. We specifically focus on fully-connected layers since they are usually the largest in terms of size in a typical DNN model, and their execution is fundamentally more memory-bounded than other types of layers (e.g. convolutional layers). In addition, a wide set of popular DNN architectures are entirely composed by FC layers, such as LSTMs and Transformers, highly relevant for time-series and natural language processing tasks. Moreover, multilayer perceptrons (MLPs) are already the status quo in use cases with very tight resource constraints, since many studies identified MLPs to be one of the best algorithms to solve tasks in the IoT domain using wearable devices \cite{wang_fann_mcu_2020}. We apply several optimization techniques from both, the hardware and software fronts, all tailored to increase the area efficiency and lower the power consumption of inference. Our goal is ultimately to make state-of-the-art MLP models more amenable for, e.g.,  the aforementioned applications.

Our contributions can be summarized as follows:
\begin{itemize}
    \item Firstly, we design a specialized hardware accelerator, named \textit{FantastIC4}, which implements a first-accumulate-then-multiply computational paradigm (ACM) in order to minimize the required number of multipliers for inference down to only 4 (thus the name of the architecture). By implementing ACM we significantly reduce the computational resource utilization compared to the usual multiply-accumulate (MAC) paradigm, due to (1) performing less number of multiplication in total, (2) enabling a more optimal data movement of the activations for MLP models (activation stationary), and (3) reducing the required area and power consumption for computations. 
    
    \item FantastIC4 also supports the efficient, on-chip execution of multiple compressed representations of the weight parameters of FC layers. This boosts the compression rate of the layers, consequently improving the off- and on-chip data movement, thus saving in power consumption as well as area requirements since lower-sized memory units can be implemented.
    
    \item In order to make the models amenable for the efficient execution on FantastIC4, we propose a novel training algorithm that makes the models robust to 4bit quantization while simultaneously encouraging low entropy statistics of the weights. Explicitly enforcing low entropy statistics reduces the size-requirements of the parameters and encourages sparsity simultaneously, which we exploit by converting the parameters to compressed sparse formats.

    \item Our experimental results show that we can save 80\% energy by compression and avoid unwanted data movement between the DDR3 DRAM and the on-chip SRAM and 75\% of power by handling the 4bit precision and sparsity in the processing elements (PEs).
    
    \item We evaluate the FantastIC4 on FC layers of popular DNN models, as well as on custom multilayer perceptrons (MLPs) trained on hand-gesture and speech recognition tasks. We compare our accelerator to other state-of-the-art FPGA and ASIC accelerators, and see an improvement by  51$\times$ in terms  of  throughput  and by 145$\times$ in  terms  of  area  efficiency (GOPS/mm$^2$).
\end{itemize}

In section~\ref{sec: related work} we describe the other state-of-the-art techniques both on the hardware and software platform. Section~\ref{sec: rationale} we describe the need for using 4-bit quantization and how we handle the sparsity. The complete hardware architecture with PE design and other floating point operations is described in section~\ref{sec: hardware}. In section~\ref{software} we explain the training of the 4bit-compact DNNs. The experimental methodology is explained in section~\ref{sec: experiments}, followed by the conclusion in section~\ref{conclusion}.
\section{Related work}
\label{sec: related work}
In recent years there has been a plethora of work published on the topic of efficient processing of DNNs, ranging from topics of neural architecture search, pruning or sparsification, quantization, compression and designing specialized hardware architectures. \cite{VS_survey, SH_survey} give an excellent overview on the landscape of different approaches and techniques studied in this topic. 

\subsection{Techniques for reducing the information content of the DNNs parameters}
The work presented in \cite{DeepCompression} pioneered a particular compression paradigm for DNN models that is based on chaining sparsification, quantization and lossless compression methods together, attaining up to 49$\times$ compression ratios. However, several follow up works have been able to achieve improvements on all three fronts.

\textbf{Lossless compression}.
The authors in \cite{DeepCABAC} showed that by coupling quantization with a powerful universal entropy coder, the compression gains can be boosted to 63$\times$ on the same models. Although the proposed method achieves impressive compression gains, the resulting representation of the DNNs weights requires decoding in order to perform inference. In contrast, similar to the \textit{Compressed Sparse Row} (CSR) matrix format employed in \cite{DeepCompression}, \cite{CER} derives a representation that compresses the weights and enables inference in the compressed representation without requiring decoding. \cite{CER} showed that their proposed \textit{Compressed Entropy Row} (CER) matrix format is up to 2$\times$ more compact and efficient than the CSR format when applied to DNNs.

\textbf{Quantization}.
In recent years researchers have been able to push more and more the limits of quantization. In particular, there is a growing corpora of work showing that extreme quantization of the weights down to 4bits is possible, while minimally affecting the prediction accuracy of the network \cite{Bhalgat_2020_CVPR_Workshops, pwlq, shkolnik_robust_2020, Jung_2019_CVPR, Gong_2019_ICCV, yang_searching_2020, zhong_towards_2020}. 4bit quantization offers directly 8$\times$ compression gains and similar improvements in computational efficiency. Stronger quantization techniques such as ternary and binary networks have also been proposed \cite{XNOR_net, EC2T}. Although they offer highly efficient implementations on a hardware level, they usually come at the expense of significant degradation of the accuracy of the network.

\textbf{Simultaneous optimization of sparsity, quantization and compression}.
Some recent work have attempted to derive a unified framework for sparsifying, quantizing and compressing  DNNs parameters. In particular, some have proposed novel regularizers that constrain the entropy of the weight parameters during training, thus explicitly minimizing the information content of the weights \cite{ECT, EC2T, ECT_google}. Concretely, in these works the first-order entropy is considered, that is, the entropy value as measured by the empirical probability mass distribution  of the parameters. This regularization technique is theoretically well motivated, directly measures the possible size reduction of the model and encourages sparsity and quantization of the weights to low bit-widths simultaneously. These works were able to attain state-of-the-art compression results, e.g., \cite{EC2T} was able to train highly sparse and ternary DNNs, becoming one of the top 5 finalists in the NeurIPS 2019 Micronet Challenge\footnote{\url{https://micronet-challenge.github.io}}.

\subsection{Hardware accelerators}

There are large number of hardware accelerators from both the academia and industry that are concentrating on high performance as well as energy efficiency. Some of the topics that have been studied and analyzed are:

\textbf{Data Flow Movement}. Data flow movement is one of the key aspects in designing the hardware accelerators for any AI application. Effective movement of the weights and activations help in reducing a large amount of energy and the power requirement. The work in \cite{Chen2017} provides an effective row stationary method and competent reusing of weights, input feature maps (Ifmaps) and partial sums (Psums) reuse. 
The Psums truncation from each of the preceding layers and performing inference on the truncated Psums and weights was shown in \cite{suhas2020}. Bit Fusion \cite{Sharma2018} dynamically shared the weights across the different layers of a DNN model. The FantastIC4 concentrates on reducing the data movement by 4bit precision and using FIFOs as a data buffer. Bitmask encoding to fetch the data from the FIFOs based on the sparsity. In addition, FantastIC4 also supports effective handling of layer weights by fetching the bit mask encoded non-zero values in a FIFO manner. Lastly, the floating point operations are pipelined to ensure the dynamic power is saved without compromising on the accuracy.


\textbf{Sparse Data Compression}. The compression with sparsity and pruning was shown in \cite{DeepCompression} to fit the DNN models in the on-chip SRAM. Based on sparsity, the hardware accelerator is implemented in \cite{Han2016} and it is 19$\times$\ more energy efficient than the uncompressed versions. The compression was further extended to convolutional layers in \cite{Parashar2017}. The weights and activations were compressed using CSC format \cite{eyerissv2}. The scalpel accelerator \cite{Yu2017} showed that the weight pruning achieves a total speedup of 1.9$\times$. In contrast to FantastIC4, all mentioned accelerators support only one particular compressed format which can greatly limit the attainable compression gains and consequently the power savings from off-chip to on-chip data movement.

\textbf{FPGA based Accelerators}. A number of FPGA accelerators have proposed solutions for optimized accelerator designs both in the industry and academia. The energy-efficient FPGA accelerator in \cite{Duan2019b} performed inference on CNN with binary weights. The processor achieves a throughput of 2100 GOPs with a latency of 4.6ms and power of 28W. The hardware-software co-design library to efficiently accelerate the entire CNN and FCN on FPGAs was shown in \cite{Zhang2019}. The floating point arithmetic CNN accelerator \cite{Lian2019a} introduced an optimized quantization scheme based on rounding and shifting-operations, they reported an overall throughput of 760.83 GOPs. Other accelerators worked on sparse matrix-vector multiplications, applied mainly to multilayer perceptrons \cite{Wang2018, Shi2019}. Even though these accelerators have a good performance, they still lack either in throughput, power or latency requirements. The FantastIC4 FPGA version, utilizes an efficient computation approach to achieve high throughput, with minimal power, latency and resource requirements.

\section{Rationale behind FantastIC4's design}
\label{sec: rationale}

\begin{figure*}[t]
    \centering
    \includegraphics[width=0.85\textwidth]{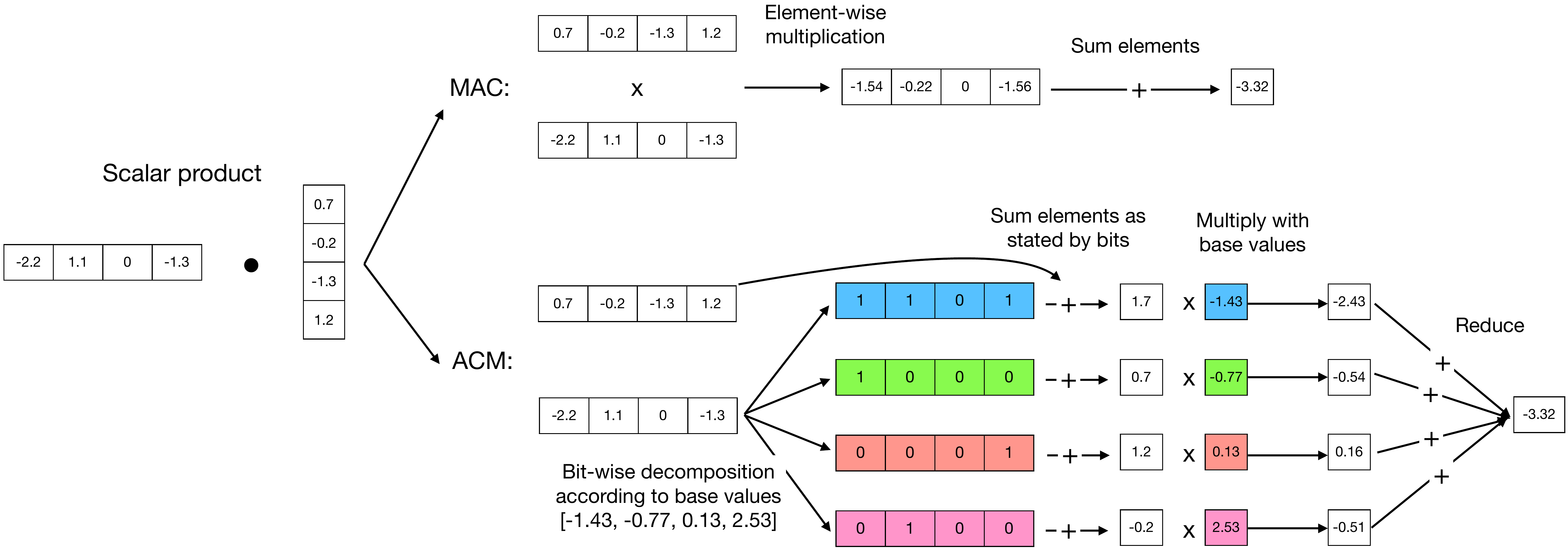}
    \caption{Sketch example on the different computational paradigms when performing the dot product algorithm. Given two input vectors, the multiply-accumulate (MAC) calculates the respective scalar product by firstly multiplying the elements and subsequently adding them. In contrast, the accumulate-multiply (ACM) firstly sums the elements of one of the vectors (in this diagram the right-hand-side vector) according to the bit-decomposition of the other, then multiplies the respective basis values and finally reduces the output. In the above sketch the base values were [-1.43, -0.77, 0.13, 2.53], and we color-coded according to [blue, green, red, pink] respectively. Thus, the original element values result by performing the linear combination in the vertical direction, for instance, $-2.2 = 1\times (-1.43) + 1\times (-0.77) + 0\times (0.13) + 0\times (2.53)$.}
    \label{Fig: MAC_ACM}
\end{figure*}

In this work we propose to apply several optimization techniques that,  in combination,  are tailored to reduce both,  area and energy requirements for performing inference. The main idea is to apply techniques that minimize the memory requirements as well as the number of multiplications needed to perform inference, since both are the major source of area utilization and power consumption.

\subsection{Why do we focus on 4bit quantization?}
As mentioned in the related work (section \ref{sec: related work}), it is well known that quantization is a powerful technique for lowering the memory as well as computational resources for inference \cite{VS_survey, SH_survey}. The increasing demand for deployment of DNNs on edge devices with very tight hardware constraints (e.g. microcontrollers) has pushed researchers to investigate methods for extreme quantization, resulting in weights with merely 4bits or lower. This directly translates to 8$\times$ compression of the model, which is beneficial for minimizing the costs involved in off- and on-chip data movement of the weights.  In particular, FC layers have shown to be highly redundant and robust to extreme quantizations down to 4bit \cite{DeepCompression, Han2016}, which again is the main focus of our work. 

\subsubsection{(Contribution 1) Increasing the computational efficiency}
However, most often the inference modules of extremely quantized layers are implemented following the usual multiply-accumulate (MAC) computational paradigm as shown in fig.~\ref{Fig: MAC_ACM}. We argue that in the regime of extreme low precision this computational paradigm is not the most efficient. Instead, we propose to first accumulate the activations at each bit-level and subsequently multiply the results, thus an accumulate-multiply (ACM) computational paradigm. More concretely, we follow the equation
\begin{equation}
    \underbrace{W \cdot A}_{\text{MAC}}  = \left(\sum_{i=0}^{3} \omega_i B_i\right) \cdot A= \underbrace{\sum_{i=0}^{3} \omega_i (B_i \cdot A)}_{\text{ACM}}
    \label{Eq: ACM}
\end{equation}
where we denote as $W$ the weight parameters of, e.g., a fully-connected layer, $A$ the input activations, $\cdot$ the operator denoting the dot product and $B_i$ a binary mask corresponding to the base $\omega_i$. Thus, as shown in equation \eqref{Eq: ACM}, we represent the weight parameters $W$ as a linear combination of four binary masks $B_i$ with respective coefficients $\omega_i$. This representation generalizes any type of 4bit-representation that is applied to the weights. For instance, if $\omega_i=2^i$ then the elements of $W$ are simply represented in the uint4 format.

As one can see from the right-hand side of equation~\eqref{Eq: ACM}, we can first accumulate the activation values that are positioned as indicated by the bitmasks $B_i$, and then multiply the output by the base value (or base centroid) $\omega_i$. This significantly reduces the required number of multiplications. Concretely, in our setup only 4 multiplications are required per output element, which is almost negligible for large dimensions of the input activations. Thus, the inference procedure is dominated by the complexity of performing additions.

\subsubsection{(Contribution 2) Mixed-precision elements}
\begin{figure}[t]
    \centering
    \includegraphics[scale=0.6]{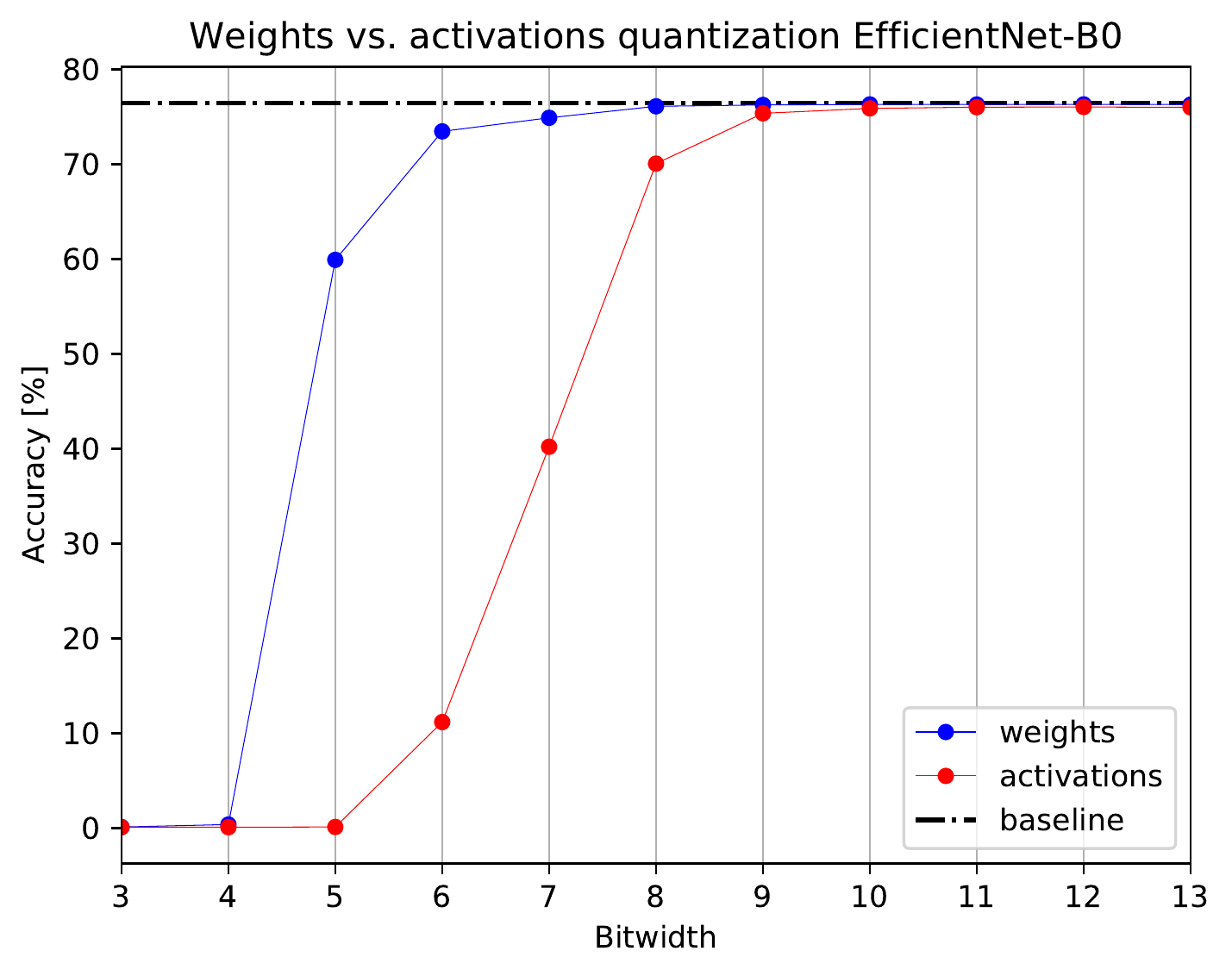}
    \caption{Difference in sensitivity between the activations and weight parameters of the EfficientNet-B0 model. Activations are more sensitive to quantization since the prediction performance of the model drops significantly faster (at higher precision values).}
    \label{Fig: w_vs_act_sensitivity}
\end{figure}

Activations are often more sensitive to perturbations than the weights as shown in fig.~\ref{Fig: w_vs_act_sensitivity}. Moreover, special parameters such as bias and batch-normalization tend to also be more sensitive than the weight parameters. This motivates the support of mixed-precision layers where input and output activations, as well as bias and batch-norm parameters can be represented with higher precision than the weights in order to compensate for the accuracy degradation. FantastIC4s design supports higher precision activation values, since this can be easily integrated within the ACM computational flow. In addition, we support full-precision representation of the batch-norm parameters as well as the bias coefficients, since their memory and compute cost are relatively low as compared to the operations involved in the weight parameters. 

In addition, in our work we do not constrain the linear coefficients values $\omega_i$ to be of powers of 2, as it is most common in the MAC approach, but allow $\omega_i \in \mathbb{R}$. This increases the expressive power of $W$, and with it the capacity of the model, allowing it to better learn more complex tasks (section \ref{sec: experiments}).

\subsection{Why do we focus on low entropy?}
As thoroughly discussed in \cite{CER}, lowering the entropy of the weights comes with a series of benefits in terms of memory as well as computational complexity. We stress that by entropy we mean the first-order entropy, that is, as measured by the empirical probability mass distribution  of the parameters. Concretely, $H = -\sum_i P_i\log_2{P_i}$, where $P_i$ measures the empirical probability mass distribution of the $i$-th cluster center. In the following we explain how in this work we leverage on the low-entropy statistics of the weights.

\subsubsection{(Contribution 3) Saving arithmetic operations}
Low entropy statistics encourage sparsity \cite{CER}. As thoroughly explained in previous work \cite{Han2016, Chen2017, Parashar2017}, sparsity allows to save computations by skipping zero-valued operations. In particular, FantastIC4 does not perform additions of activations when zero-valued weights are present, thus saving on arithmetic operations and consequently dynamic energy consumption.

Moreover, low entropy statistics do also encourage low number of unique non-zero values, thus a high probability of encountering the same non-zero value. This property can be exploited when loading non-zero values, by reducing the dynamic power required when loading the same value. 

\subsubsection{(Contribution 4) Multiple lossless compression} 
There are several ways to compress sparse weights. One is by converting the weights in the \textit{Compressed Sparse Row} (CSR) format \cite{DeepCompression}, which is based on applying run-length coding for saving the signaling of the positions of non-zero values. Another one is by applying a simple form of Huffman coding, which consists of storing a bitmask indicating the positions of the non-zero values followed by an array of non-zero values organized in, e.g., row-major order. In the high sparsity regime ($>$90\% of zeros), the CSR format attains higher compression gains, whereas for smaller sparsity ratios (25\% - 90\% of zeros) the Huffman code compresses more the weights. Since the sparsity ratio of different layers can vary significantly, FantastIC4 supports the processing of both sparse representations on-chip. This allows for more flexible compression opportunities, consequently boosting the compression gains of the model and saving on off- to on-chip transmission costs.

\section{Training 4bit-compact DNNs}
\label{software}

\begin{figure*}[t]
    \centering
    \includegraphics[width=\textwidth]{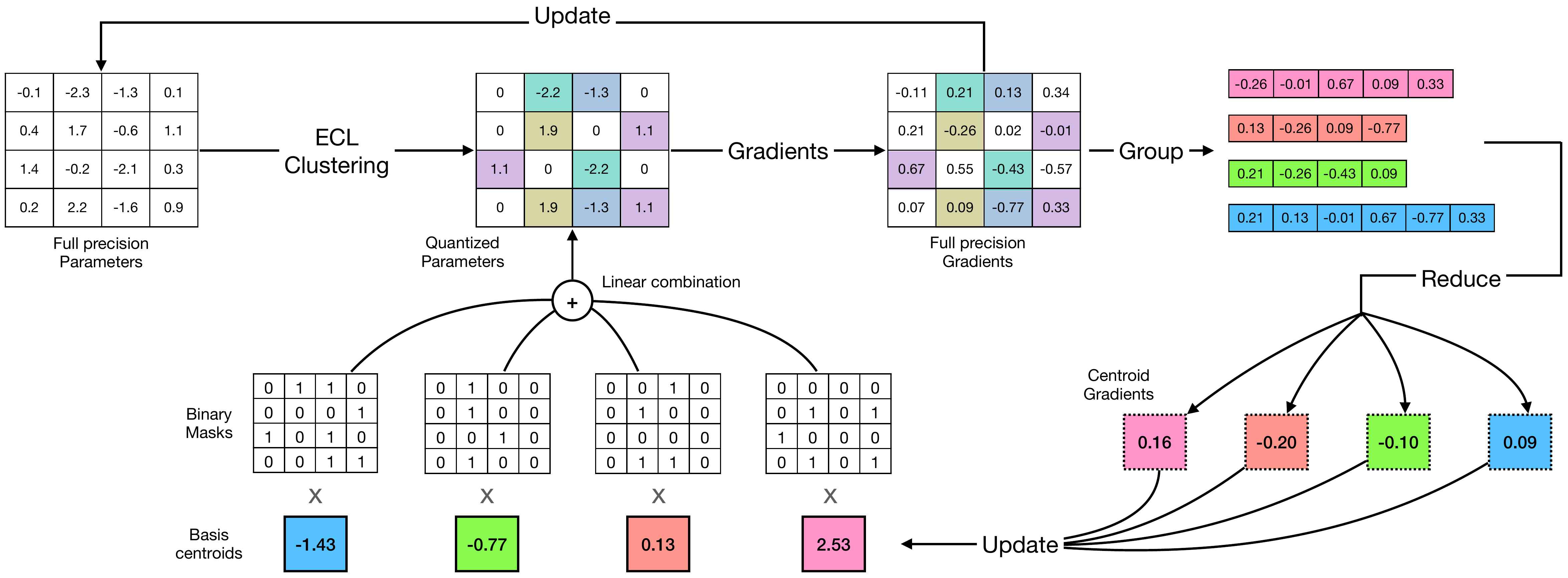}
    \caption{4bit-entropy-constrained training method for compressing DNNs, based on the straight through estimator (STE). Firstly, the full-precision parameters are quantized using the entropy-constrained Lloyd (ECL) algorithm, whereas the quantization points are constrained to be linear combinations of 4 bitmasks with 4 basis centroids. Then, the gradients are calculated w.r.t. the quantized DNN model. The full-precision parameters are respectively updated, whereas the gradients of each basis centroid are computed by grouping and reducing their respective shared gradient values.}
    \label{Fig: STE_ECL}
\end{figure*}

As described in the previous Sections \ref{sec: rationale}, our proposed optimization paradigm is based on the fact that the weight parameters exhibit low-entropy statistics and can be represented with 4bits. However, if we naively lower the entropy and strongly quantize a pretrained model then, most often, we would incur a significant drop in accuracy (see experimental section \ref{sec: experiments}). Therefore, in this work we propose a novel training algorithm that makes DNN models robust to such type of transformations. 

\subsection{Entropy-constrained training of DNNs}
Our method is strongly based on EC2T, a method proposed in \cite{EC2T} that trains sparse and ternary DNNs to state-of-the-art accuracies. We generalize their approach so that DNNs with 4bit weights and low entropy statistics are attained instead. Concretely, our training algorithm is composed by the following steps:
\begin{enumerate}
    \item Quantize the weight parameters (but keep a copy of the full-precision weights) by applying the entropy-constrained Lloyd (ECL) algorithm \cite{source_coding}.
    \item Apply the straight-through estimator (STE) \cite{STE} and forward + backward pass the quantized version of the model.
    \item Update the full-precision weights and the centroids with the computed gradients.
\end{enumerate}
Fig. \ref{Fig: STE_ECL} sketches the training method.

\subsection{Definition of the centroids} 
As described in equation \eqref{Eq: ACM} (section \ref{sec: rationale}), we represent the weight parameters $W$ of the DNN as a linear combination of 4 binary masks $B_i$ with respective coefficients $\omega_i$. This allows us to define 16 different cluster center values (or centroids), with four of them being the coefficients $\omega_i$ and the rest a particular linear combination of them. In order to increase the capacity of the models, we assign to each weight parameter $W$ their unique set of four centroids $\Omega$. 

\subsection{Entropy-Constrained Lloyd algorithm (ECL)} 
The ECL algorithm is a clustering algorithm that also takes the entropy of the weight distributions into account. We stress that throughout this work we define entropy as $H = -\sum_i P_i\log_2{P_i}$, where $P_i$ measures the empirical probability mass distribution of the $i$-th cluster center. To recall, the $H$ states the minimum average amount of bits required to store the output samples of the distribution \cite{Shannon}. Thus, ECL tries not to only minimize the distance between the centroids and the parameter values, but also the information content of the clusters. Again, this regularization term is theoretically well motivated, directly measures the possible size reduction of the model and encourages sparsity + quantization of the weights to low bit-widths.

However, we slightly modify the algorithm so that the cluster centers are not updated by the ECL method. Instead, we fine-tune the cluster centers with the information received from the gradients (more in the later subsection \ref{subsec: fine-tuning}). 

\subsection{Making DNNs robust to post-training quantization} 
As we stated earlier, if we naively apply the ECL algorithm to a pretrained network, then the accuracy drop may be significant. Therefore, we apply the STE method \cite{STE} in order to make them robust to extreme quantization. In the case of NNs this simply means to apply further training iterations where we update the the full-precision parameters with regards to the gradients computed by the quantized parameters. By doing so we adapt the full-precision weight parameters to the prediction error incurred by the quantization, thus forcing them to move to minima where they are robust to ECL-based quantization.

\subsection{Fine-tuning centroids} 
\label{subsec: fine-tuning}

Our particular contribution is reflected in the definition of the 16 clusters and their respective gradient propagation (i.e. fine-tuning). To recall, we represent each (quantized) weight parameter as a linear combination of 4 binary masks $B_i$ with respective coefficients $\omega_i$, thus $W = \sum_{i=0}^3 \omega_i B_i$. Therefore, we only update the four basis centroids $\omega_i$ at each training iteration, since 12 out of the 16 centroids are linear combinations of these.  Hence, we calculate the gradients $\delta^{\omega}_i$ of each centroid $\omega_i$ as follows: Let $\delta^{W}$ be the gradient tensor of the weight parameter $W$, then
\begin{equation}
    \delta^{\omega}_i = \sum_{j=0} \delta^{W}_j B_{i}
\label{Eq: centroid gradient}
\end{equation}
with $B_{i}$ being the binary mask respective to the coefficient $\omega_i$, and $j$ being the dimension that iterates over all parameter elements. 

After computing the gradient of each centroid, we update them by applying the ADAM optimizer \cite{ADAM}.

\section{Fantastic4: Specialized hardware accelerator for running 4bit-compact DNNs}
\label{sec: hardware}
\begin{figure}[t!]
    \centering
    \includegraphics[trim=0.5cm 0.5cm 0.5cm 0.5cm,clip=true,width=0.4\textwidth]{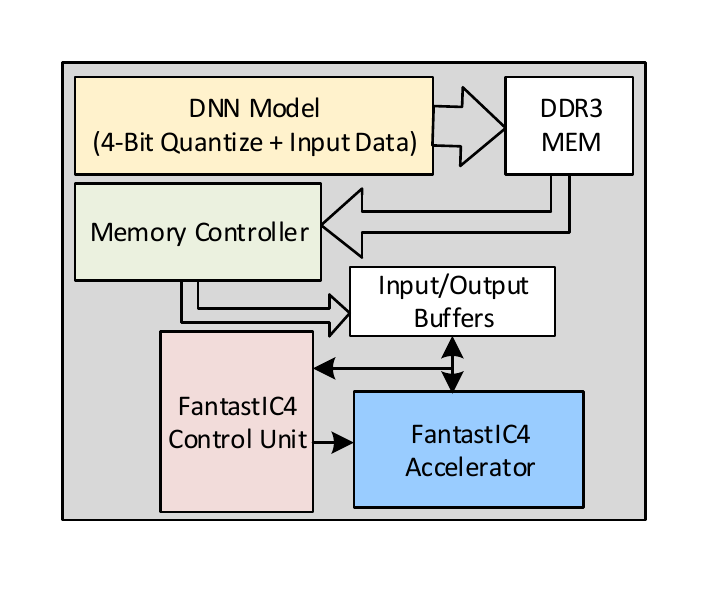}
    \caption{FantastIC4 System.}
    \label{Fig: FantastIC4_sys}
\end{figure}

\begin{table*}[t!]
  \caption{Control States of the FantastIC4 Control Unit}
  \label{tab:Control}
  \resizebox{2\columnwidth}{!}{
    \begin{tabular}{|c|>{\columncolor[HTML]{FFFC9E}}c |c|>{\columncolor[HTML]{DAE8FC}}c |c|c|c|c|c|c|c|}
    \hline
    \textbf{\makecell{Data \\ Movement}} & - &
  \cellcolor[HTML]{DAE8FC}\makecell{Acts,Wt,Bias\\ alpha and CSR} &
  CSR Data &
  \cellcolor[HTML]{FFFC9E}- &
  \cellcolor[HTML]{FFFC9E}- &
  \cellcolor[HTML]{FFFC9E}- &
  \cellcolor[HTML]{FFFC9E}- &
  \cellcolor[HTML]{FFFC9E}- &
  \cellcolor[HTML]{FFFC9E}- &
  \cellcolor[HTML]{FFFC9E}- \\ \hline
\textbf{Computation} &
  - &
  \cellcolor[HTML]{FFFC9E}- &
  BM Conv &
  \cellcolor[HTML]{DAE8FC}Wt ID &
  \cellcolor[HTML]{DAE8FC}\begin{tabular}[c]{@{}c@{}}Add tree/\\ MAC\end{tabular} &
  \cellcolor[HTML]{DAE8FC}Fix-Flt &
  \cellcolor[HTML]{DAE8FC}\begin{tabular}[c]{@{}c@{}}FLT\\ Mul1\end{tabular} &
  \cellcolor[HTML]{DAE8FC}\begin{tabular}[c]{@{}c@{}}Flt \\ Add\end{tabular} &
  \cellcolor[HTML]{DAE8FC}\begin{tabular}[c]{@{}c@{}}Float \\ Mul1\end{tabular} &
  \cellcolor[HTML]{DAE8FC}Float-Int \\ \hline
\textbf{State} &
  \cellcolor[HTML]{FFCCC9}Start &
  \cellcolor[HTML]{FFCCC9}State1 &
  \cellcolor[HTML]{FFCCC9}State2 &
  \cellcolor[HTML]{FFCCC9}State3 &
  \cellcolor[HTML]{FFCCC9}State4 &
  \cellcolor[HTML]{FFCCC9}State5 &
  \cellcolor[HTML]{FFCCC9}State6 &
  \cellcolor[HTML]{FFCCC9}State7 &
  \cellcolor[HTML]{FFCCC9}State8 &
  \cellcolor[HTML]{FFCCC9}State9 \\ \hline
\textbf{Time(ns)} &
  \cellcolor[HTML]{34FF34}0 &
  \cellcolor[HTML]{34FF34}5000 &
  \cellcolor[HTML]{34FF34}10 &
  \cellcolor[HTML]{34FF34}10 &
  \cellcolor[HTML]{34FF34}10 &
  \cellcolor[HTML]{34FF34}30 &
  \cellcolor[HTML]{34FF34}50 &
  \cellcolor[HTML]{34FF34}50 &
  \cellcolor[HTML]{34FF34}40 &
  \cellcolor[HTML]{34FF34}20 \\ \hline
\end{tabular}
}
\end{table*}

\begin{figure*}[t!]
    \centering
    \includegraphics[trim=0.5cm 0.5cm 0.5cm 0.5cm,clip=true,width=0.85\textwidth]{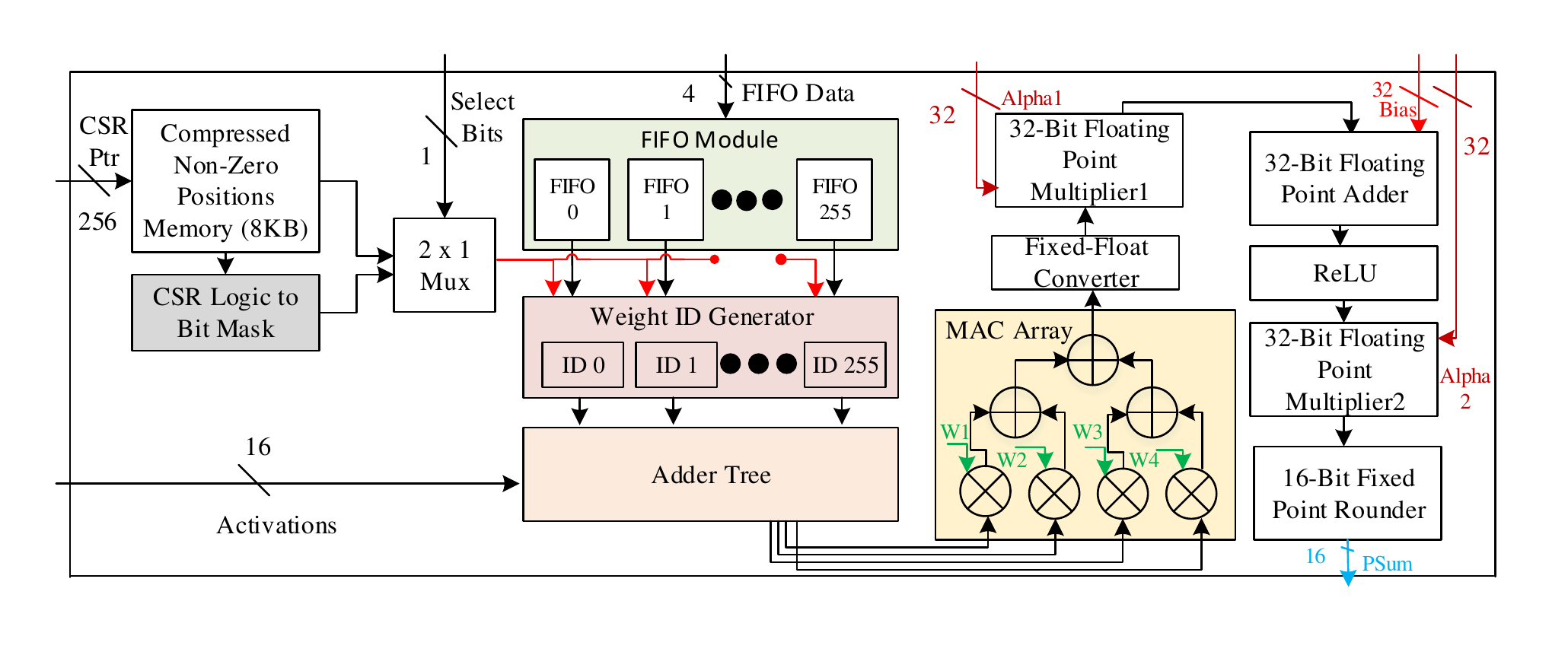}
    \caption{FantastIC4 Architecture.}
    \label{Fig: FantastIC4_Arch}
\end{figure*}

Fig.~\ref{Fig: FantastIC4_sys} shows the overview of the FantastIC4 system. The whole system is a heterogeneous combination of a CPU and an FPGA architecture. The entire system comprises of mainly three parts: the software program on the CPU, the external DDR3 memory and the hardware architecture on the FPGA chip. The software part mainly consists of the CPU that transfers the input data as well as the DNN model (only one time) to the FPGA. Since all the data is usually very large and can therefore not entirely be stored on an on-chip BRAM, some of it is stored in an off-chip DRAM. The data is then accessed through a memory controller which is built across a memory interface generator (MIG) IP. On the FPGA chip, we have the FantastIC4 control unit, memory controller, I/O Buffers and the FantastIC4 accelerator. The memory controller facilitates the movement of the input data from off-chip DRAM to the accelerator and stores back the computation results into the DRAM. The control unit regulates the behaviour of other modules on the FPGA, it handles the data movement and the computation inside the accelerator. The I/O buffers store the input data for processing and store back the PSum data from the accelerator for inference of the subsequent layer. The FantastIC4 accelerator is the heart of the entire system which reads the data from the DRAM, performs the computation and stores back the results into the DRAM memory.

\subsection{Memory Controller and Input/Output Buffers}
The DDR3 memory is accessed by the FantastIC4 accelerator through a MIG interface operating at a clock frequency of 200MHz. We employ the AXI communication protocol for the data movement between the FPGA chip and the off-chip DRAM. The microblaze CPU and other AXI control IPs are used to communicate through the MIG interface with the DDR3. The memory controller receives instructions from the FantastIC4 control unit through the AXI master to read and write the data from/to the  memory. The I/O buffers provide the dual buffering for data movement in a ping-pong manner.

\subsection{FantastIC4 Control unit}
Our proposed accelerator has two levels of control hierarchy. Table~\ref{tab:Control} shows the control states for our accelerator. The first level of hierarchy, i.e. Start and State1, control the data movement between the DRAM, memory controller and the accelerator on the FPGA chip. Here the activations, weights, biases, alpha values for floating point operations, FIFO data and 256bits CSR Pointer data are moved into their respective memory/registers for computation. In this level all the data movement operations are performed sequentially, the total time taken to complete these two states are approximately around 5000ns for MLP models. Here the total time taken is mainly dependent on the DNN model which is under inference. In the next level of hierarchy we perform the computations, State2-State9 shows the different stages of processing performed on the accelerator. The different orders of computation performed are: CSR to bitmask conversion, weight ID generation, accumulation and multiply operation and finally the single precision floating point operation. The total time taken to perform the computation is around 220ns. The computation time is less because all the states are working concurrently and each state is independent on the other states except on the first iteration.

\subsection{FantastIC4 Architecture}
The top-level hierarchy of the FantastIC4 architecture is shown in Fig.~\ref{Fig: FantastIC4_Arch}. The architecture operates on a single clock frequency domain of 150MHz (FPGA Based Implementation) and 800MHz (ASIC Based Implementation). FantastIC4 is composed of CSR to bit mask logic to perform CSR to bit mask conversion, FIFO modules to store the weight IDs for 256 adder trees, weight ID generator fetches the data from the FIFO modules based on the outcome of CSR to bit mask conversion. An adder tree performs the accumulation of the activations based on the weights IDs from the ID generator. The MAC array performs four multiplication and three addition operations. The fixed point to floating point converter converts a 16bit fixed point MAC output into a 32bit single precision float output. This 32bit floating point MAC output will be multiplied by a 32bit alpha1 values; where alpha1 values are an array of single precision floating point data, the output of the multiplier1 will be added with the bias. The output of the adder will undergo a non-linear activation operation called ReLU, to perform the computation $f(x) = max(0,x)$. Final floating point multiplication is performed with another 32bit single precision alpha2 value, the 32bit result from the multiplication will be rounded back to 16bit integer value to generate the final PSum.

\begin{figure}[t]
    \centering
    \includegraphics[trim=0.75cm 0.65cm 0.5cm 0.5cm,clip=true,width=0.4\textwidth]{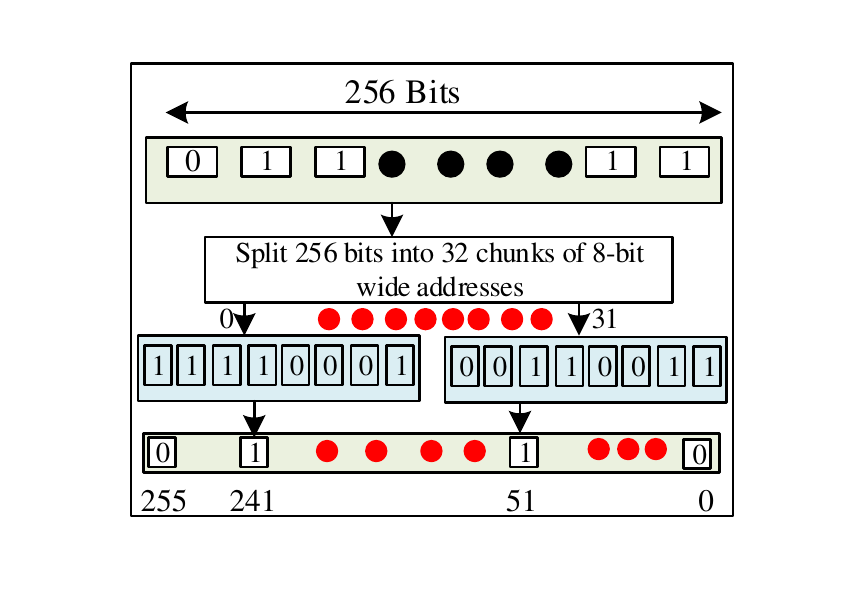}
    \caption{CSR to bitmask Conversion Logic.}
    \label{Fig: CSR2BM}
\end{figure}

\textbf{CSR to bitmask Logic.} 
By default, FantastIC4 loads the positions of the non-zero elements of a row of the sparse weight matrix according to the compressed Huffman representation, which consists of a simple binary mask of width 256. The bitmask controls the weight ID movement into the adder tree. However, when a layers non-zero positions are compressed following the CSR format, a logic must be implemented that converts them back to a bitmask representation, which is the purpose of the CSR to bitmask Logic. The conversion logic is shown in Fig.~\ref{Fig: CSR2BM}, the compressed non-zero position data pointers comprising of 256 bits will be splitted into a chunks of 32 which is of 8bit wide. Based on the 8bit value, each bit of the encoded bitmask will be set to `1'. For ex: As shown in Fig.~\ref{Fig: CSR2BM}, the 0\textsuperscript{th} chunk had a value of 241 and 31\textsuperscript{st} chunk had a value of 51. So the corresponding 241\textsuperscript{st} bit and 51\textsuperscript{st} bit will be set to 1 and the remaining bits will be set to 0 to generate a 256bits encoded bitmask data. Both the CSR pointer data and bitmask data will be selected from a 2 $\times$ 1 Mux through a ``Select Bits'' to generate the final encoded data for weight ID generation.

\textbf{FIFO Module and Weight ID Generator.}
The FIFO module has 256 individual FIFOs which has a width of 4 and depth of 256, each storing the non-zero weight elements of a particular column of the weight matrix of the layer. These FIFO modules are stored in array of registers. The weight ID generator has a simple selection logic, where each individual IDs of 4bits are fetched from the FIFOs based on the encoded bitmask data. If the encoded bitmask is `1', then an ID will be fetched from the FIFOs or else the pointer points to the same location of the fetched data. The weight ID generator has a cluster of 256 ID modules, which store the 4bit IDs from the FIFO if the 256 individual bits from the bitmask is `1' or else it stores a 4bit zero data.

\begin{figure}[t]
    \centering
    \includegraphics[trim=0.75cm 0.65cm 0.5cm 0.5cm,clip=true,width=0.5\textwidth]{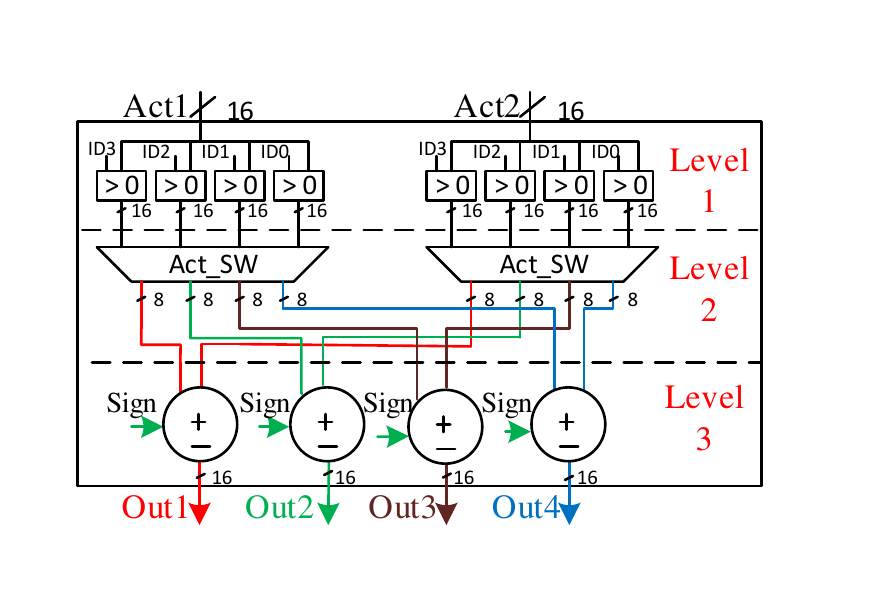}
    \caption{Adder Schematic.}
    \label{Fig:Adder_Sch}
\end{figure}

\textbf{Adder Tree and MAC Array.}
An adder tree comprises of an array of 256 adders arranged in a logarithmic fashion. The adder tree is grouped into two stages: Adder Stage1 and Adder Stage2. Adder Stage1 has 128 adders arranged in a single group.  Each adder in the stage1 has three levels of hierarchy with a control parameter in each hierarchy. The Fig.~\ref{Fig:Adder_Sch} shows the adder schematic in the Stage1. The adders are fed with the two-different activations and two-different IDs from the weight ID generator. All the activations in the adder tree are static and it is used for all cycles of computations. The static activations in the adder tree saves significant power consumption. By having a static activations inside the adder rather than accessing it from the memory saves up to 15\% of power consumption. In the level1 hierarchy, the 4bit weight IDs control the movement of the activations inside the adder. Each bit from the weight IDs forms a channel, that regulate the flow of activation to the level2. If the ID is 1, then the 16bit activation is fed or else a zero value is fed to level2. There will be a total of eight groups of activation data coming out of the level1 hierarchy. In level2 hierarchy, the activation switches between the upper and lower half of 8bits. This technique is employed to fit the larger networks into the hardware and improvise the prediction in the hardware. In this hierarchy, if the activation switch is low, a lower half of the bits are selected or else the  upper half is selected. In the level3 hierarchy, the actual computation is performed. The sign mode determines, whether the activations need to be added or subtracted. Finally the four different computations are performed among the eight groups,  to generate the four output data of 16bits from each adder. The Adder Stage2 has 128 adders arranged in a multiple group. The first group in adder stage2 has 64 adders, second group has 32 adders and similarly other groups are scaled down logarithmically. The adders in adder stage2 performs only the computation, unlike the adders in stage1. Based on the sign-bit in the output data from the stage1, either addition or subtraction is performed.

The MAC array performs four multiplications and three additions respectively. The four outputs each of 16bits from the adder tree will be multiplied with the 16bit basis weights to generate a 32bit product, which we will be accumulated to generate the final 32bit MAC output.

\textbf{Floating Point Operations.}
The floating point operation mainly comprises of fixed to floating point conversion, floating point multiplications, floating point addition and final 32bit floating point to 16bit integer conversion. In the fixed to floating point conversion, a 32bit fixed point MAC data is converted into equivalent single precision floating point data as shown in Algorithm.~\ref{alg:Fixed Point to Floating Point Conversion}, a leading one will be detected from the MAC output and corresponding conversion operation is performed.

\begin{figure}[t]
    \centering
    \includegraphics[trim=0.75cm 0.65cm 0.5cm 0.5cm,clip=true,width=0.4\textwidth]{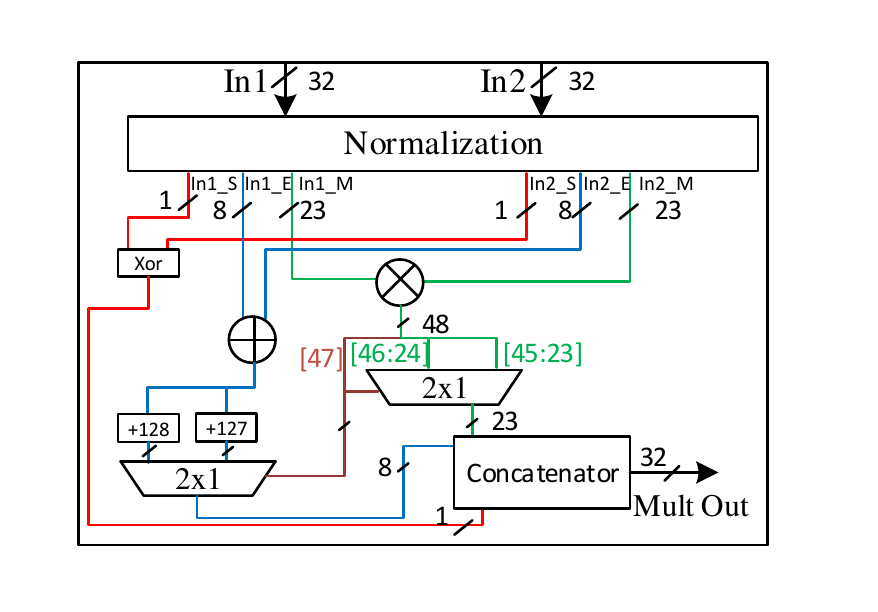}
    \caption{Floating Point Multiplier.}
    \label{Fig:Float_Mult}
\end{figure}

The converted floating point data will undergo a single precision floating point multiplication with Alpha1 values. The Alpha1 values are stored in a SRAM of 1KB. With this scaling factors FantastIC4 is able to accommodate for de-quantization as well as batchnorm parameters.  As shown in Fig.~\ref{Fig:Float_Mult}, both the inputs will be normalized and split into it equivalent sign, mantissa and exponent part. The 23bits mantissa will be multiplied with each other to generate 48bit output, the MSB of the multiplied output will be used to calculate the final mantissa and the exponent part.The sign bits of both the inputs will be XORed to generate the final sign bit. The final sign, exponent and mantissa will be concatenated to generate the final 32bits multiplied output. 
Subsequently, the multiplied floating point will be  added with the bias data stored in another 1KB SRAM. This operation is similar to the multiplication operation in terms of normalization of the data. Then, the added data will undergo an nonlinear activation ReLU function as $f(x) = max(0,x)$, since it is the status quo non-linear function for most MLP models. The ReLUed output will be further multiplied with a single 32bit Alpha2 value to generate the final 32bit multiplied output. These scaling factors take further quantization parameters into account, important for the correct calibration of the subsequent quantization step, which consists of a final rounding of 32bits to a 16bit integer. The 16bit integer is the final PSum that will be used as an activations for the inference of the next layer.

\begin{algorithm}[t]
\caption{Fixed Point to Floating Point Conversion}
\label{alg:Fixed Point to Floating Point Conversion}
\begin{algorithmic}
\State $mac\_out$: 32bit fixed point MAC output 
\State $convert\_out$: Single precision floating point number
    \Procedure{FixedtoFloat}{$mac\_out$, $convert\_out$}
    \State $lod \gets$ Leading one detector
    \State $convert\_out\_sign \gets$ Sign bit of floating point
    \State $convert\_out\_exponent \gets$ Exponent of floating point
    \State $convert\_out\_mantissa \gets$ Mantissa of floating point
    \For{$k \gets 30$ downto $0$} 
        \If{($mac\_out$[k] == 1)}
            \State $lod$ = k;
        \EndIf
    \EndFor
    \State $convert\_out\_sign$ = $mac\_out[31]$;
    \State $convert\_out\_exponent$ = $lod$ + $127$;
    \State $convert\_out\_mantissa$ = $mac\_out$ \verb|<<| ($23$ - $lod$);
    \State Combine sign,exponent and mantissa to generate convert\_out
    \State \textbf{return} $convert\_out$
    \EndProcedure
\end{algorithmic}
\end{algorithm}

\section{Experiments}
\label{sec: experiments}

\subsection{Experimental setup}
In the experiments section we distinguish between hardware-conform and non-conform models. Conform models are those that are fully compatible with our hardware architecture, thus the entire end-to-end inference procedure can be performed on it. Consequently, conform models include only FC layers with up to 512 input/output features. Optionally, BatchNorm layers are allowed which can result in accuracy gains.

\subsubsection{Datasets \& Models}
To cover a variety of use-cases with the conform models, we trained and deployed several models solving classification tasks for audio, image and biomedical data. Concretely, we considered the task of hand-gesture recognition (HR) based on biomedical and sensor data, the Google Speech Commands (GSC) dataset for the task of audio classification, MNIST and CIFAR-10 datasets for small-scale image classification task. We trained and implemented custom and well-known MLPs for solving the above mentioned tasks. In addition, in order to benchmark our quantization algorithm we also used non-conform models, which we have not trained ourselves but obtained from publicly available sources. Concretely, ResNet-50 and -34 come from the torchvision model zoo\footnote{\label{torchvision}\url{https://pytorch.org/docs/stable/torchvision/models.html}}, EfficientNet-B0 from \footnote{\label{effnetb0}\url{https://github.com/lukemelas/EfficientNet-PyTorch}, Apache License, Version 2.0 - Copyright (c) 2019 Luke Melas-Kyriazi} and ResNet-20 from \footnote{\label{res20}\url{https://github.com/akamaster/pytorch_resnet_cifar10}, Yerlan Idelbayev's ResNet implementation for CIFAR10/CIFAR100 in PyTorch}. We trained further these models by applying our entropy-constrained method (section \ref{software}), and benchmarked their accuracies at different regularization strengths.
$\quad$\\
\textbf{Hand Gesture Recognition (HR).}
The authors in \cite{georgi_recognizing_2015} collected Inertial Measurement Unit (IMU) and electromyogram (EMG) readings from 5 different subjects in 5 different sessions in order to capture 12 defined hand gestures. Different from \cite{georgi_recognizing_2015}, we deploy a small MLP to solve the classification task. It clearly outperforms the proposed Hidden Markov Model which achieves a mean accuracy of 74.3$\%$ for person-independent hand gesture recognition. Our proposed 4-layer deep MLP achieves a person-independent mean accuracy of 84.0$\%$. Quantizing all network layers to 4 bit with the FantastIC4 algorithm is possible with almost no drop in accuracy.
The model consists of an input layer, two hidden layers and an output layer with 512, 256, 128 and 12 output features, where a BatchNorm layer follows each fully connected layer. The data corpus is publicly available \footnote{\label{handgesture}\url{https://www.uni-bremen.de/en/csl/research/motion-recognition.html}}.

\textbf{Google Speech Commands (GSC).} The Google Speech Commands dataset consists of 105,829 utterances of 35 words recorded from 2,618 speakers. The standard is to discriminate ten words "Yes", "No", "Up", "Down", "Left", "Right", "On", "Off", "Stop", and "Go", and adding two additional labels, one for “Unknown Words”, and another for “Silence” (no speech detected) \cite{warden_speech_2018}. There are no overlapping speakers between the train, test and validation sets. We deploy a MLP consisting of an input layer, five hidden layers and an output layer featuring 512, 512, 256, 256, 128, 128 and 12 output features, respectively. Our model achieves a classification accuracy of 91.0$\%$ which outperforms the default CNN model (88.2$\%$) in the TensorFlow example code mentioned in \cite{warden_speech_2018}. The FantastIC4 4 bit quantization has a regularizing effect on the full-precision MLP and further improves the classification accuracy to 91.35$\%$, while introducing ~60$\%$ sparsity. The authors in \cite{zhang_hello_2018} show that for the GSC dataset CNNs and especially RNNs usually achieve better accuracies than MLPs. Still, our proposed model yields a comparable accuracy to their proposed CNN and outperforms their 8 bit quantized MLP (88.91$\%$). For another comparison, \cite{liu_ultra_low_2019} quantized their network composed by three convolution layers and two fully-connected layers to 7 bit using 8 bit activations, and achieve an accuracy of 90.82$\%$.

\textbf{Image Classification.} For small-scale image classification we utilized two neural networks, one MLP which would fit into our proposed accelerator, LeNet-300-100, and one CNN (ResNet-20). CIFAR-10 \cite{krizhevsky_learning_2009} is a dataset consisting of natural images with a resolution of $32\times 32$ pixels. It contains 10 classes. The train and test sets contain 50,000 and 10,000 images. MNIST \cite{lecun_mnist_2010} is drawn from 10 classes where each class refers to a handwritten digit (0-9). The dataset contains 60,000 training images and 10,000 test images with a resolution of $28\times 28$ pixels. To benchmark our quantization algorithm with ImageNet we deployed EfficientNet-B0, ResNet-50, and -34 networks. The ImageNet \cite{deng_imagenet_2009} dataset is a large-scale dataset containing 1.2 million training images and 50,000 test images of 1000 classes. The resolution of the image data is various and in the range of several hundred pixels. We crop the ImageNet data in all experiments to $224\times 224$ pixels.

\subsubsection{Hardware simulation setup}
The proposed FantastIC4 was implemented in System Verilog and corresponding behavioral and gate level simulation was performed using Mentor Graphics Simulator. The FPGA version of the FantastIC4 was implemented using Xilinx Vivado tool. Here we synthesized, place and routed the design on a Virtex Ultrascale FPGA on the XCVU440 device. 

For the ASIC version, we synthesized the architecture using Synopsys Design Compiler (DC) under the GF 22nm FDSOI SLVT technology. We placed and routed the design using Synopsys IC compiler (ICC2). After the sign-off and RC extraction using STARRC, we performed the timing closure using Synopsys Prime-Time. We annotated the toggle rates from the gate level simulation and dumped the toggling information into Value Change Dump (VCD) file and estimated the power using Prime-Time.

\subsection{Benchmarking training of 4bit-compact DNNs}
\begin{figure}[t]
    \centering
    \includegraphics[scale=0.6]{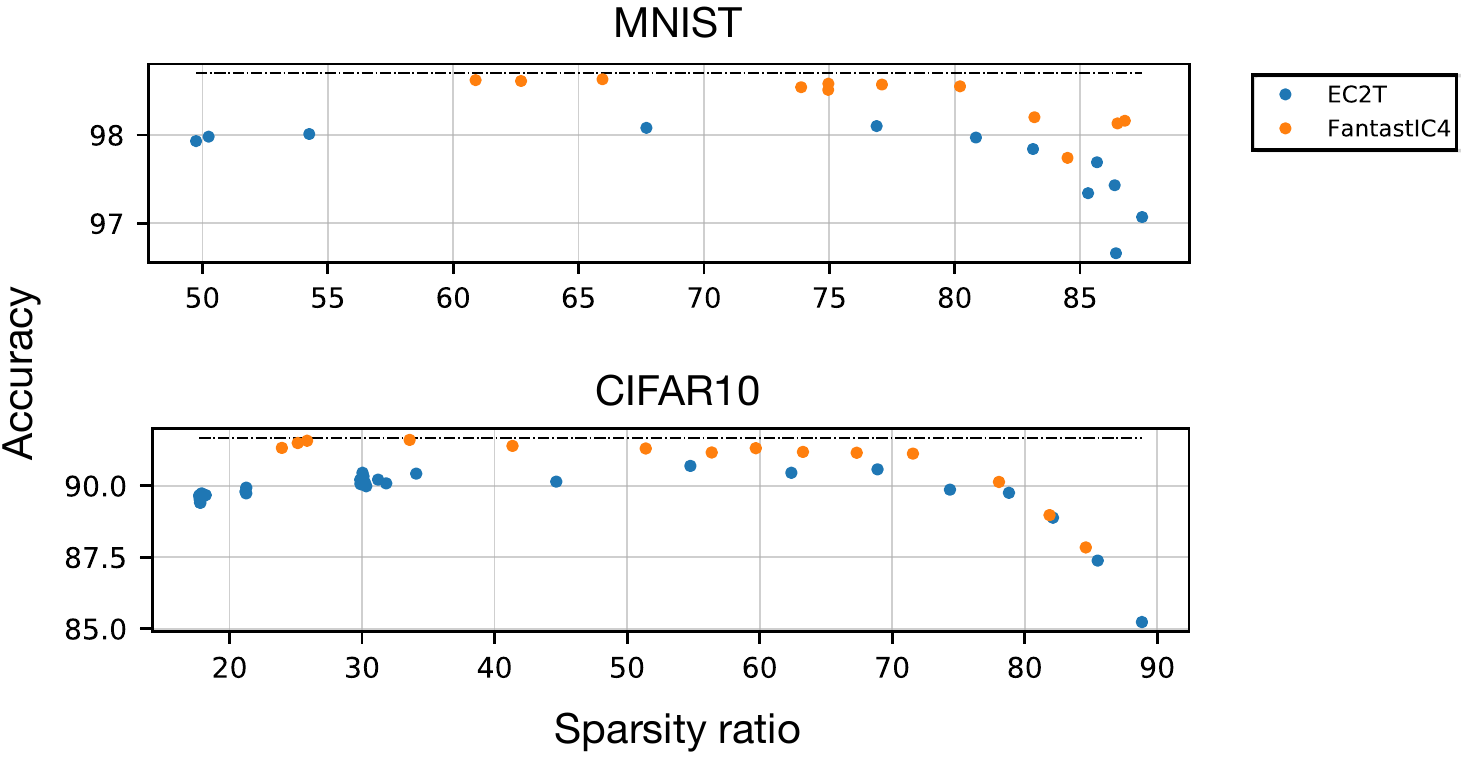}
    \caption{Accuracy as a function of the sparsity ratio of different DNN models. (Top) LeNet-300-100 model trained on the MNIST dataset by the previous method EC2T \cite{EC2T}, as compared to FantastIC4s generalized form of entropy-constrained training method. (Bottom) same as top, but for ResNet20 trained on the CIFAR10 dataset.}
    \label{Fig: EC2T vs Fanta4}
\end{figure}

Arguably, the closest related training method to ours is EC2T, which trains sparse and ternary networks under an entropy-constrained regularizer \cite{EC2T}. However, our generalization allows to train DNNs with more expressive power due to their ability to express 16 different cluster centers instead of only 3. Moreover, thanks to the support of full-precision scaling factors which can accommodate for batchnorm parameters we expect our DNN models to be more robust to strong quantization + sparsification, consequently attaining better prediction performance vs compression trade-offs. Fig. \ref{Fig: EC2T vs Fanta4} shows this phenomena. We can see that our DNN models trained with our entropy-constrained approach reach better Pareto-optimal fronts with regards to accuracy vs sparsity, as compared to the EC2T method.

Furthermore, in Table \ref{table: F4_vs_other} we show the prediction performance of our models on the datasets described above and summarize the results attained by other authors. We can see that we consistently attain similar or higher prediction performance than the previous work. Table \ref{table: F4_vs_other} also shows the benefits of applying a hybrid compression scheme as opposed to the single compression format approach as proposed by previous work. To recall, our compression scheme encodes each layer by applying the CSR, the simple Huffman code (or bitmask format), and the trivial 4bit dense representation, and chooses the most compact representation between them. We see, that we attain about 2.36 $\times$ boost in compression gains on average as compared to the CSR-only approach proposed by \cite{Han2016, eyerissv2}, and $\times$1.77 higher compression rates than the trivial 4bit dense format. These gains directly translate to reduction in memory, off- to on-chip data movement and area requirements, which stresses the importance of supporting multiple representations.

\begin{table}[!th]
\renewcommand{\arraystretch}{1.0}
\caption{Comparison of the FantastIC4-quantization approach vs previous state-of-the-art 4bit quantization techniques. For each network we report two results, one showing highest accuracies we attained and the other highest compression ratios. Our models as well as best results are highlighted in bold. All models belonging to the same row-block have the same architecture, with exception of the Google Speech Command and Hand Gesture Recognition datasets. Unless otherwise specified, all approaches quantize all network layers, including input- and output-layers, excluding batch normalization- and bias-parameters.}
\label{table: F4_vs_other}	
\begin{center}
	\begin{tabular}{llcccclll}
		\toprule
		\bfseries Model & \bfseries Org. Acc.~($\%$) &  \bfseries Acc.  & \bfseries Size (MB) & \bfseries CR$^a$ & \bfseries CSR$^b$ \\
		\midrule
		\multicolumn{6}{c}{\bfseries ImageNet} \\
		\midrule
        \bfseries EfficientNet-B0 & 76.43 &   \bfseries 75.01 & 21.15 & 7.62 & 3.31 \\
         \bfseries EfficientNet-B0  & 76.43 & 74.08  & 21.15 & \bfseries 8.25 & 3.91 \\
        LSQ+~\cite{Bhalgat_2020_CVPR_Workshops} & 76.10 & 73.80 & 21.15 & 7.48 & 2.59 \\
        \midrule
        \bfseries ResNet-50 & 76.15  &  \bfseries 75.66 & 102.23 & 8.21 & 3.50 \\
         \bfseries ResNet-50 & 76.15 & 75.29 & 102.23 & \bfseries 9.97 & 4.50 \\
        PWLQ~\cite{pwlq}$^\dagger$ & 76.13 & 75.62 & 102.23 & 7.86 & 2.64 & \\
        KURE~\cite{shkolnik_robust_2020}$^\dagger$ & 76.30 & 75.60 & 102.23 & 7.88 & 2.64\\
        \midrule
        \bfseries ResNet-34$_{IO}^\dagger$ & 73.30  &  72.98 & 87.19 &  7.80 & 4.32 \\
        \bfseries ResNet-34$_{I}^\dagger$ & 73.30 & 72.86 & 87.19 & \bfseries 9.30 & 4.37 \\
        QIL~\cite{Jung_2019_CVPR}$^\dagger$ & 73.70 & \bfseries 73.70 & 87.19 & 6.82 & 2.65 \\
        DSQ~\cite{Gong_2019_ICCV}$^\dagger$ & 73.80 & 72.76 & 87.19 & 6.82 & 2.65 \\
		\midrule
		\multicolumn{6}{c}{\bfseries CIFAR-10} \\
		\midrule
		\bfseries ResNet-20 &  91.67 & 91.60 & 1.08 & 8.43 & 3.92 \\
		\bfseries ResNet-20 & 91.67 & 91.15 & 1.08 & \bfseries 16.23 & 11.31 \\
		SLB~\cite{yang_searching_2020}$^\dagger$ & 92.10 & \bfseries 92.10 & 1.08 & 7.64 & 2.62 \\
		GWS~\cite{zhong_towards_2020}$^\dagger$ & 92.20 & 91.46 & 1.08 & 7.72 & 2.62 \\
		\midrule
		\multicolumn{6}{c}{\bfseries MNIST} \\
		\midrule
		\bfseries LeNet-300-100  & 98.70 & \bfseries 98.63  & 1.07 & 13.31 & 7.62 \\
		\bfseries LeNet-300-100 & 98.70 & 98.16 & 1.07 & \bfseries 29.31 & 19.81 \\
		\midrule
		\multicolumn{6}{c}{\bfseries Google Speech Commands} \\
		\midrule
		\bfseries MLP-GSC & 91.00  & \bfseries 91.19 & 2.57 & 10.88 & 5.55 \\
		\bfseries MLP-GSC & 91.00 & 90.41 & 2.57 & \bfseries 13.59 & 7.99 \\
		 HE~\cite{HE} & 86.40 & 86.40 & 0.2 & - & - \\
		\midrule
		\multicolumn{6}{c}{\bfseries Hand Gesture Recognition} \\
		\midrule
		\bfseries MLP-HR & 88.50 & \bfseries 88.33 & 1.30 & 8.51 & 3.96 \\
	     \bfseries MLP-HR  & 88.50 & 87.22 & 1.30 & \bfseries 13.57 & 8.35 \\
	     HMM\cite{georgi_recognizing_2015} & 74.30 & 74.30 & - & - & - \\
		\bottomrule
	\end{tabular}\\
	\footnotesize{$^{a}$ Compression ratio defined as the ratio of the full-precision model size to the quantized model size, where FantastIC4 stores each layer in its optimal format which is either CSR, bitmask format or the trivial 4bit dense format.} \\
	\footnotesize{$^{b}$ Compression ratio defined as the ratio of the full-precision model size to the quantized model size, where each layer is stored in CSR format.} \\
	\footnotesize{$^\dagger$ QIL and DSQ use full-precision (32bit) for the first and last layer, PWLQ and SLB use full-precision for the first layer and KURE and GWS provide no information about first/last layer quantization. Our ResNet-34$_{IO}$ benchmark has 32bit input- and output-layers and the ResNet-34$_{I}$ benchmark a 32bit input layer.} 
\end{center}
\end{table}

\subsection{Benchmarking hardware efficiency}

\begin{table*}[t!]
 \caption{FantastIC4 Resource Utilization Breakdown for Different DNN Models on a Virtex Ultrascale FPGA. Here BR stands for BRAMs, FF for Flip Flops, LUT for Look Up Tables, DSP for Digital Signal Processing and LR stands for LUTRAMs}
 \label{tab:Resource Utilization Breakdown Summary}
 \resizebox{2\columnwidth}{!}{
\begin{tabular}{|c|c|c|c|c|c|c|c|c|c|c|c|c|c|c|c|c|c|c|c|c|c|c|c|c|c|}
\hline
\multirow{2}{*}{Modules} &
  \multicolumn{5}{c|}{MLP-HR} &
  \multicolumn{5}{c|}{EfficientNet-B0} &
  \multicolumn{5}{c|}{MobileNet-V3} &
  \multicolumn{5}{c|}{ResNet-50} &
  \multicolumn{5}{c|}{MLP-GSC} \\ \cline{2-26} 
           & LUT & FF  & BR & DSP & LR & LUT  & FF   & BR & DSP & LR   & LUT  & FF   & BR & DSP & LR   & LUT   & FF   & BR & DSP & LR   & LUT & FF  & BR & DSP & LR \\ \hline
CSR to BM  & 5K  & 255 & 0  & 0   & 0  & 5K   & 255  & 0  & 0   & 0    & 5K   & 255  & 0  & 0   & 0    & 5K    & 255  & 0  & 0   & 0    & 5K  & 255 & 0  & 0   & 0  \\ \hline
Wt ID Gen  & 0   & 512 & 0  & 0   & 0  & 0    & 512  & 0  & 0   & 0    & 0    & 512  & 0  & 0   & 0    & 0     & 512  & 0  & 0   & 0    & 0   & 512 & 0  & 0   & 0  \\ \hline
\begin{tabular}[c]{@{}c@{}}MAC \\ Array\end{tabular} &
  565 &
  153 &
  0 &
  4 &
  0 &
  570 &
  158 &
  0 &
  4 &
  0 &
  568 &
  156 &
  0 &
  4 &
  0 &
  580 &
  162 &
  0 &
  4 &
  0 &
  566 &
  153 &
  0 &
  4 &
  0 \\ \hline
\begin{tabular}[c]{@{}c@{}}Fixed point \\ to Float point \\ Op\end{tabular} &
  707 &
  483 &
  8 &
  4 &
  0 &
  717 &
  491 &
  8 &
  4 &
  0 &
  711 &
  17 &
  12 &
  4 &
  0 &
  719 &
  500 &
  16 &
  4 &
  0 &
  709 &
  484 &
  8 &
  4 &
  0 \\ \hline
Adder tree & 35K & 4K  & 0  & 0   & 0  & 35K  & 4K   & 0  & 0   & 0    & 35K  & 4K   & 0  & 0   & 0    & 35K   & 4K   & 0  & 0   & 0    & 35K & 4K  & 0  & 0   & 0  \\ \hline
\begin{tabular}[c]{@{}c@{}}FIFO\\  Module\end{tabular} &
  53K &
  6K &
  0 &
  0 &
  8K &
  103K &
  119K &
  0 &
  0 &
  160K &
  830K &
  95K &
  0 &
  0 &
  128K &
  1661K &
  190K &
  0 &
  0 &
  256K &
  53K &
  6K &
  0 &
  0 &
  8K \\ \hline
BM Memory  & 21  & 821 & 8  & 0   & 0  & 38   & 842  & 36 & 0   & 0    & 33   & 834  & 31 & 0   & 0    & 48    & 821  & 63 & 0   & 0    & 28  & 822 & 8  & 0   & 0  \\ \hline
Total      & 95K & 12K & 16 & 8   & 8K & 108K & 125K & 44 & 8   & 160K & 872K & 101K & 43 & 8   & 128K & 1703K & 196K & 79 & 8   & 256K & 95K & 12K & 16 & 8   & 8K \\ \hline
\end{tabular}
}
\end{table*}

\subsubsection{Results on MLPs}

\begin{table}[t!]
\caption{FantastIC4 Final Resource Utilization.}
  \label{tab:Final Resource Utilization}
   \resizebox{1\columnwidth}{!}{
\begin{tabular}{|c|c|c|c|c|c|}
\hline
Resource   & LUT     & LUTRAM  & FF      & BRAMs  & DSP  \\ \hline
Used       & 1703,187  & 128,000  & 196,909   & 79     & 8    \\ \hline
Avaliable  & 2532,960 & 459,360  & 5065,920 & 2,520   & 2,880 \\ \hline
Utlization & 67.24\% & 27.86\% & 3.88\%  & 3.13\% & 0.27\%  \\ \hline
\end{tabular}
}
\end{table}

We benchmarked FantastIC4 hardware performance on the fully-connected layers of several popular models such as  EfficientNet-B0, MobileNet-v3 \& ResNet-50. Furthermore, we benchmarked the end-to-end inference efficiency of two of our custom and fully hardware-conform multilayer perceptrons (MLPs), trained for the task of google speech commands and hand-gesture recognition. Both MLP  models, which we named \textit{MLP-GSC} \& \textit{MLP-HR} respectively, reach state-of-the-art prediction performance on their tasks (see Table \ref{table: F4_vs_other}.

Table~\ref{tab:Resource Utilization Breakdown Summary} shows the resource utilization breakdown of our FantastIC4 accelerator for different DNN models. The proclaimed results are based on the post-implementation results from Xilinx Vivado 2018.2. The activations values are quantized down to a 8bit precision, whereas the four basis weights use a precision of 16bits. This configuration was found to be accurate enough to perform the inference without harming the prediction performance of the models. As shown in Table~\ref{tab:Resource Utilization Breakdown Summary}, we consume the lowest resources among all the accelerators reported so far that perform on fully connected layers. Here we engage both the fixed point and floating point operations for faster processing and improved accuracy during the inference. We consume a total of just 8 DSPs to perform the computation which significantly reduces the dynamic power consumption. Moreover, very few BRAMs are used in the entire inference operation due to the extreme quantization and compression, the LUTRAMs are explicitly used for storing the weightIDs inside the FIFOs. As one can see, the floating point operations utilize the lowest resources on the FPGA chip due to the enhanced data flow modelling. The final resource utilization summary is shown in Table~\ref{tab:Final Resource Utilization}.

\begin{table}[t!]
\caption{Layout results of the ASIC version.}
\begin{center}
 \resizebox{0.80\columnwidth}{!}{
\begin{tabular}{|c|c|}
\hline
Technology           & GF 22nm FDSOI SLVT                           \\ \hline
Chip Size & 1mm $\times$ 1.2mm                                 \\ \hline
Core Area             & 800$\mu m$ $\times$ 800$\mu m$                        \\ \hline
Core Voltage          & 0.88V                                  \\ \hline
Memory Type          & SRAM (10KB)                                  \\ \hline
Total Gate Count     & 961K                                         \\ \hline
Frequency            & 800MHz                                      \\ \hline
Precision             & Fixed 16bit                                 \\ \hline
Power                 & 454mW                                        \\ \hline
Latency               & 1.31$\mu$s                                       \\ \hline
Performance           & 9.158 TOPS                                    \\ \hline
Performance/W         & 20.17 TOPS/W                                  \\ \hline
Energy                & 595 nJ                                       \\ \hline
DNN Models Inferenced  & MLP-HR and MLP-GSC \\ \hline
\end{tabular}
}
\end{center}
\label{tab:ASIC_Results}
\end{table}

\begin{figure}[t!]
    \centering
    \begin{tabular}{cc}
    \includegraphics[trim=3.75cm 3.25cm 0.5cm 0.5cm,clip=true,width=0.2\textwidth]{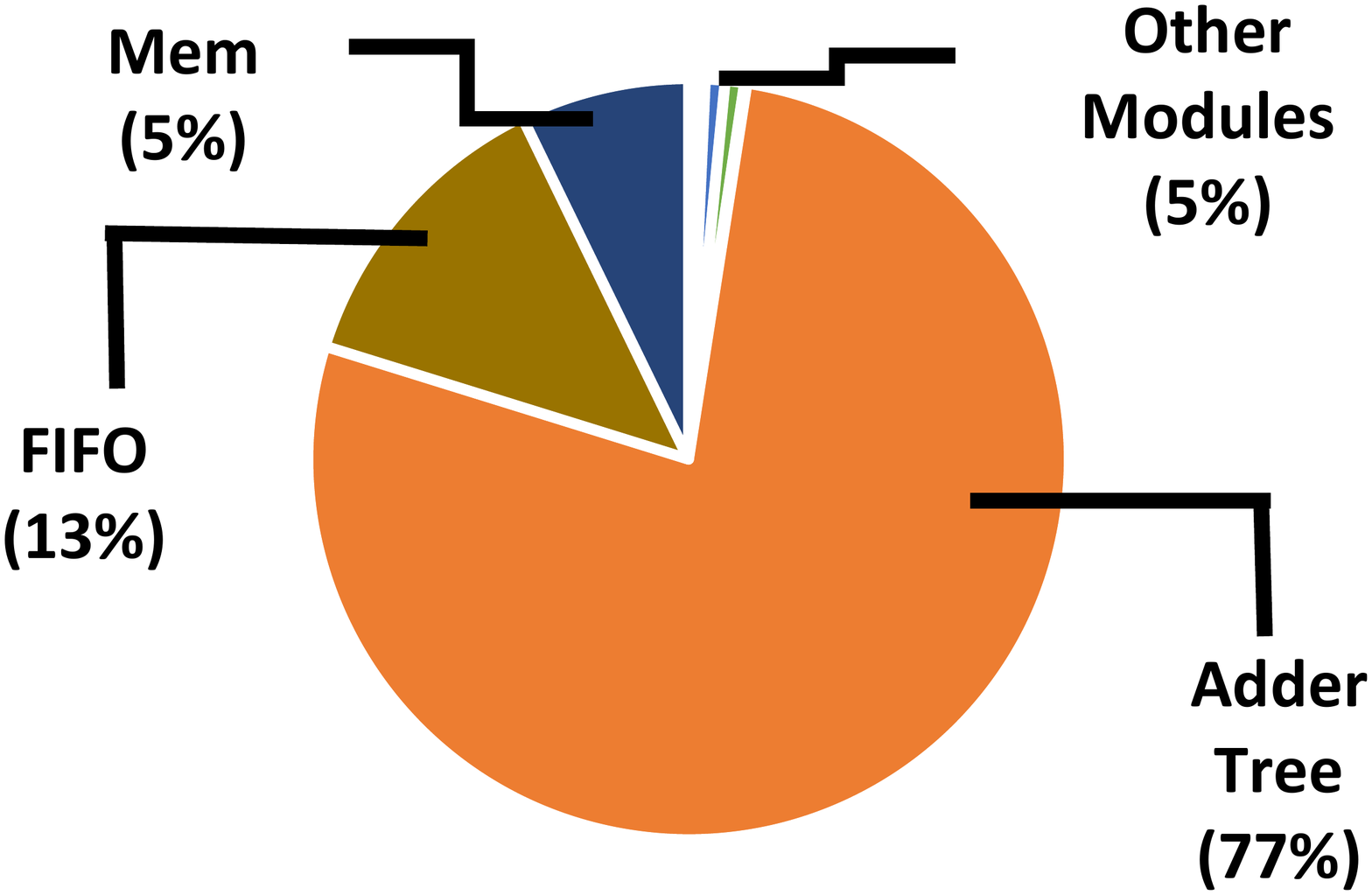} &
    \includegraphics[trim=3.75cm 3.25cm 0.5cm 0.5cm,clip=true,width=0.2\textwidth]{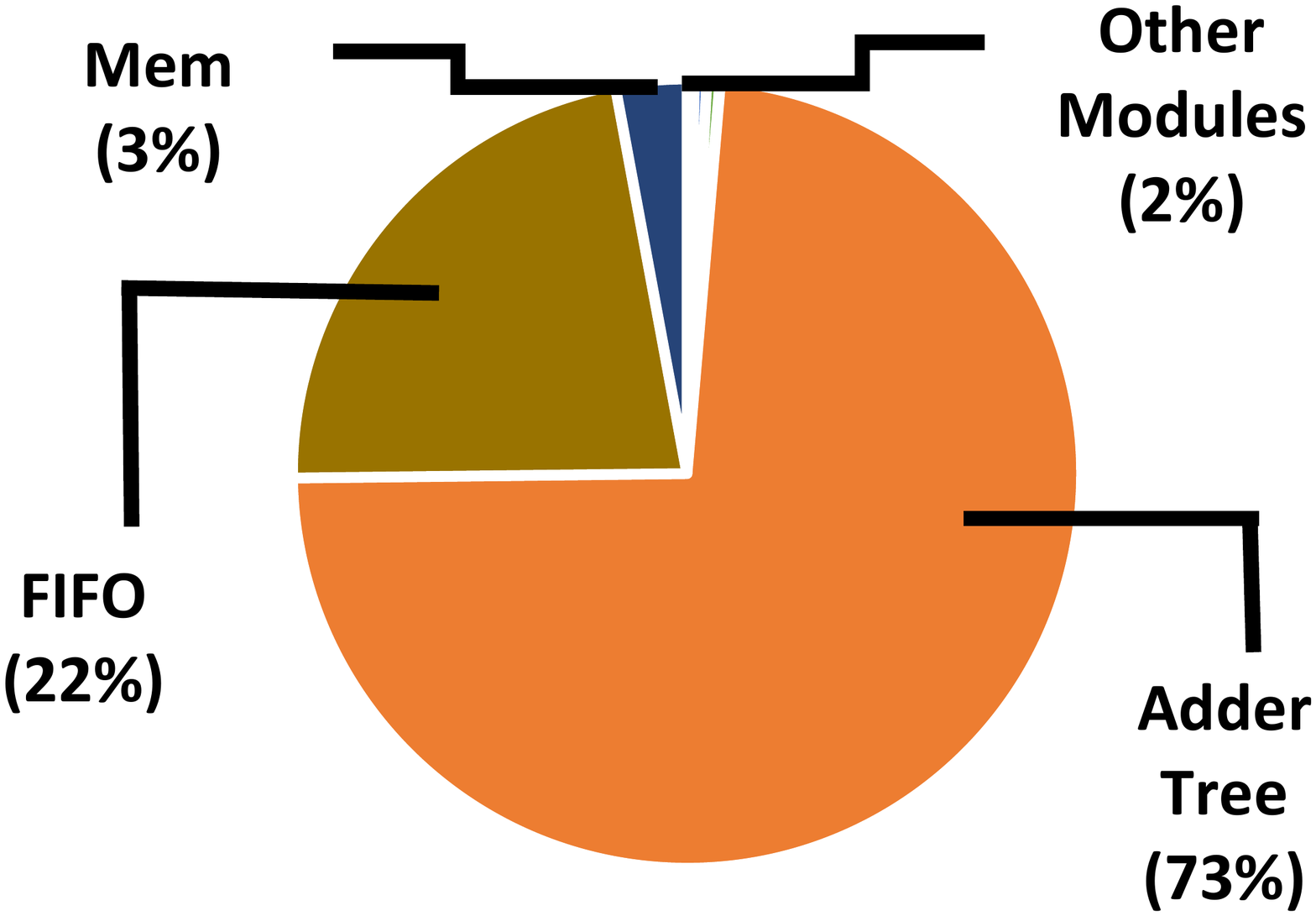} \\
    (a) Area Breakdown & (b) Power Breakdown
    \end{tabular}
    \caption{Area and Power Breakdown of FantastIC4 ASIC Version.}
    \label{Fig:Area and Power Breakdown of FantastIC4}
\end{figure}

We further evaluated the ASIC version of FantastIC4 on a 22nm process node with a clock frequency of 800MHz. The Table~\ref{tab:ASIC_Results} reports the layout version, the total area of our processor was found to be 1mm $\times$ 1.2mm. 

\subsubsection{Power Consumption, Latency and Throughput on the FPGA}
The FantastIC4 accelerator is highly energy efficient due to the low weights storage and the static activations inside the adder tree. The static activations inside the adder tree reduces the total power consumption by 15$\times$, as the reduced data movement consumed around 64mW of dynamic power when compared to the conventional SRAM access which had 960mW of power consumption.  We measured power consumption for different DNN models, throughout the inference the static power was more predominant than the dynamic power consumption, as static power on the XCVU440 FPGA was 2.856W. The total power measured from the inference of MLP-HR was 3.472W, EfficientNet-B0 was 10.14W, ResNet-50 was 12.34W, MobileNet-V3 was 8.46W and MLP-GSC was 3.6W. The average latency measurement of each layer for MLP-HR was 6.45$\mu$s, for EfficientNet-B0 was 8.6$\mu$s, MobileNet-V3 was 6.3$\mu$s, ResNet-50 was 10.2$\mu$s and MLP-GSC was 7.2$\mu$s. To infer our entire custom DNN model, we had a latency of 72$\mu$s for MLP-HR and 80$\mu$s for MLP-GSC. The overall throughput measurement was 2.45TOPS, as the processing unit remains constant irrespective of the DNN model under inference. On average, the total off-chip to on-chip data movement was saved by 10.55$\times$ as compared to the original (non-compressed) representation of the parameters. Furthermore, due to our on-chip support of hybrid compressed representations, we were able to boost the savings by 2$\times$ as compared to the compressed formats proposed by ~\cite{Han2016} and  \cite{eyerissv2}.

Similarly for the ASIC version as shown in Table~\ref{tab:ASIC_Results}, we could achieve a peak performance throughput of 13.1 TOPS and the performance/watt of 28.87 TOPS/W. The total latency to perform the inference was found to be 1.31$\mu$s for MLP-HR model and 1.37$\mu$s for MLP-GSC model. Since we first perform all the accumulation on the adder tree and then perform the MAC operation, we significantly save the resources and power required for computations by 2.7$\times$. An array of 256 MAC units with 16bit width consumes an area of  346.58$\mu$m $\times$ 346.58$\mu$m, whereas the same ACM unit will consume an area of 216.54$\mu$m $\times$ 216.54$\mu$m. Similary, an array of 256 MAC units consumes a power of 101.23mW and an array of 256 ACM units consumes a power of 40.46mW. So by our ACM technique we save  atotal area of around 39\% and power of around 40\%. The area and the power breakdown for the ASIC version is shown in the Fig.~\ref{Fig:Area and Power Breakdown of FantastIC4} Most of the area and power consumption is dominated by the adder tree and the FIFOs as it forms the core part of the architecture.

\subsection{Comparison to previous work}

\begin{table}[t!]
\caption{Performance Comparison with other State of Art FPGA Accelerators. }
\label{tab:Comparison}
\begin{center}
 \resizebox{0.80\columnwidth}{!}{
\begin{tabular}{|c|c|c|}
\hline
Parameters           & Dinelli~\cite{Dinelli2019}  & Ours     \\ \hline
Device               & XCVU65   & XCVU440  \\ \hline
Benchmark            & GSC      & GSC      \\ \hline
Quantization         & Fixed-16 & Fixed-16 \\ \hline
Sparsity             & N/A      & 60\%     \\ \hline
Accuracy             & 90.23\%      & 91\%     \\ \hline
Throughput (TOPS)     & N/A      & 2.45     \\ \hline
Throughput/W (GOPS/W) & N/A      & 198.54   \\ \hline
Static Power (W)       & 0.626    & 2.856    \\ \hline
Dynamic Power (W)     & 1.235    & 0.744    \\ \hline
Total Power (W)       & 1.861    & 3.6    \\ \hline
Latency ($\mu$s)          & 570      & 80       \\ \hline
Energy (mJ)           & 1.06     & 0.288    \\ \hline
Frequency (MHz)       & 78.4     & 150      \\ \hline
\end{tabular}
}
\end{center}
\end{table}

Here, we compare the performance of other state of art accelerator on FPGA that work on multilayer perceptrons with benchmark on the GSC dataset. The keyword spotting (KWS) accelerator~\cite{Dinelli2019} was the closest FPGA accelerator that benchmarked on google speech commands, so for fair comparison we are comparing with this accelerator. The KWS accelerator ~\cite{Dinelli2019} also quantized their DNN models and implemented the entire architecture using on-chip memories and benchmarked the results on different Xilinx and Intel FPGA devices. Table~\ref{tab:Comparison} shows the comparison results. Here we are mainly benchmarking for sparsity, accuracy, throughput and power consumption. We evaluated the performance of our accelerator on our custom MLP-GSC, as our custom built network had more sparsity and higher accuracy for KWS application.  Our FantastIC4 accelerator has an overall throughput of 2.45 TOPS due to the parallel execution of the adder tree and the MAC array and lower clock cycle requirement for the floating point operations. We have 50$\times$\ lower dynamic power consumption when we compared to  ~\cite{Dinelli2019} due to the static activations inside the adder tree, lesser number of multiplications and piplelined approach with the floating point operations. In terms of latency, we are 14$\times$\ faster to infer one complete network that works on KWS application. In terms of energy-efficiency we are 27.16$\times$\ better when compared to the other accelerator.

For the ASIC version, arguably the closest related work to FantastIC4 are EIE \cite{Han2016} \& EyerissV2 \cite{eyerissv2}, since both accelerators also leverage on compressed representations of the DNNs parameters. We stress that more recent accelerators exploiting compressed representations exist such as \cite{Parashar2017}, however, these were optimized for convolutional layers whereas FantastIC4 optimizes the execution of fully-connected layers.

In the following we provide benchmarks across different accelerators for each of those components as shown in Table~\ref{tab:Comparison_ASIC}. In the throughput comparison FantastIC4 is better than EIE by 16$\times$, Eyeriss v2 by  15$\times$ and Thinker by 31$\times$. In terms of power efficiency FantastIC4 outperforms EIE by 20$\times$, Eyeriss v2 by 14$\times$ and Thinker by 16$\times$. For the area efficiency calculation we could not compare our accelerator with Eyeriss v2 because in Eyeriss v2 area is reported in terms of total number of gates. So by comparing the total gates we are smaller by 2.9$\times$. However for other accelerators, we are better than EIE by 544$\times$ and Thinker by 406$\times$.

In Table~\ref{tab:Comparison_ASIC_KWS} we are comparing our accelerators with the other state of art ASIC KWS accelerators. Here we are comparing FantastIC4 with EERA-ASR\cite{Liu2018} and RNN based speech recognition processor \cite{Guo2019}. Both the processors work with the same Google Speech Command dataset.  We have  a better throughput by 51$\times$ and 14$\times$ when compared to other works. Similarly we are more power efficient by 6$\times$ and 1.8$\times$ respectively. In terms of area efficiency, we are efficient by 142$\times$ with respect to \cite{Liu2018} and 145$\times$ with respect to \cite{Guo2019}.

\begin{table}[t!]
\caption{Performance Comparison with other State of Art ASIC Compression Based Accelerators. }
\label{tab:Comparison_ASIC}
\begin{center}
    \resizebox{1\columnwidth}{!}{
\begin{tabular}{|c|c|c|c|c|c|c|}
\hline
Platform                  & EIE ~\cite{Han2016}     & Eyeriss V2 ~\cite{eyerissv2} & Thinker~\cite{Yin2017}   & Our's      \\ \hline
Technology (nm)           & 65       & 65         & 65          & 22       \\ \hline
Frequency (MHz)           & 800      & 200        & 200         & 800      \\ \hline
Precision                 & Fixed-16 & Fixed-16   & Fixed-8/16  & Fixed-16 \\ \hline
Throughput (GOPS)         & 572      & 858.62     & 368.4       & 9158.65  \\ \hline
Power (mW)                & 590      & 606        & 290         & 454      \\ \hline
Power Efficiency (GOPS/W) & 969.49   & 1416.87    & 1270.34     & 20173.23 \\ \hline
Area (mm\textsuperscript{2})                     & 40.8     & N/A        & 19.6        & 1.2      \\ \hline
Area Efficiency (GOPS/mm\textsuperscript{2})           & 14.02    & N/A        & 18.79       & 7632.208 \\ \hline
\end{tabular}
}
\end{center}
\end{table}

\begin{table}[t!]
\caption{Performance Comparison with other State of Art ASIC KWS Accelerators. }
\label{tab:Comparison_ASIC_KWS}
\begin{center}
\resizebox{0.9\columnwidth}{!}{
\begin{tabular}{|c|c|c|c|}
\hline
Platform                  & EERA-ASR\cite{Liu2018} & Guo\cite{Guo2019}     & Our's      \\ \hline
Technology (nm)                & 28       & 65      & 22       \\ \hline
Frequency (MHz)            & 400      & 75      & 800      \\ \hline
Latency (us)               & N/A      & 127.3   & 1.31     \\ \hline
Keywords Number           & 20       & 10      & 10       \\ \hline
Dataset                   & \multicolumn{3}{c|}{GSC}      \\ \hline
Accuracy                  & 91.88\%  & 90.20\% & 91\%     \\ \hline
Throughput (GOPS)          & 179.2    & 614.4   & 9158.65  \\ \hline
Power (mW)                 & 54       & 52.5    & 454      \\ \hline
Power Efficiency (TOPS/W) & 3.31     & 11.7    & 20.17    \\ \hline
Area(mm\textsuperscript{2})                 & 3.34     & 6.2     & 1.2      \\ \hline
Area Efficiency (GOPS/mm\textsuperscript{2}) & 53.65    & 52.51   & 7632.208 \\ \hline
\end{tabular}
}
\end{center}
\end{table}

\subsection{Ablation study: Execution efficiency of the models as a function of their entropy}
\begin{figure}[t!]
    \centering
    \includegraphics[scale=0.43]{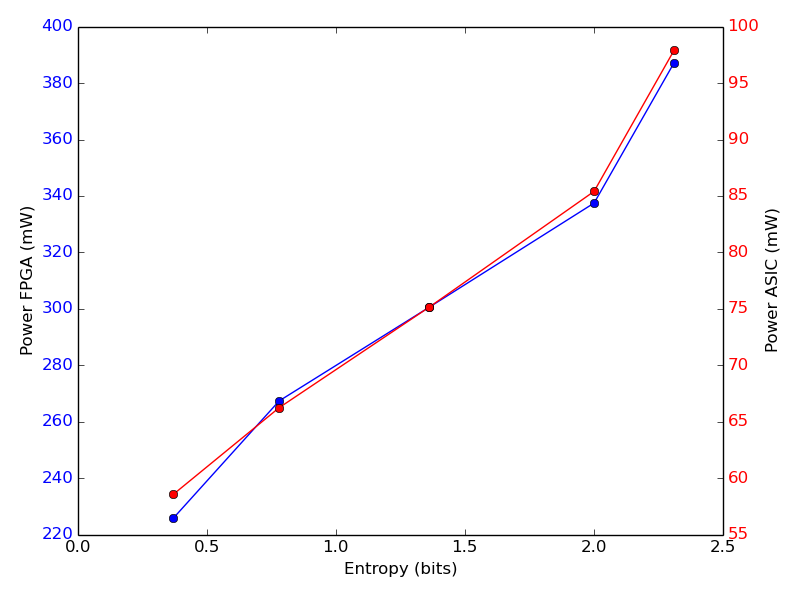}
    \caption{Power consumption of our MLP-HR model as a function of its entropy distribution. (Blue) Dynamic power consumption measured on an FPGA, (Red) measured on ASIC simulation.}
    \label{Fig: H vs power}
\end{figure}
In section III of the main manuscript we argued that one of our major contributions is the fact that FantastIC4s hardware architecture is specially designed to exploit low-entropy statistics of the weight parameters. Thus, we should expect the execution efficiency of DNN models to increase as the entropy of the weight parameters decreases. Figure \ref{Fig: H vs power} shows exactly this trend. In this experiment, we measure the power-efficiency of our MLP-HR model at different overall entropy levels of the model. To perform this study, we ran our post-layout simulation (ASIC) and post-implementation timing simulation (FPGA) to generate the corresponding Value Change Dump (VCD) for ASIC and Switching Activity Interchange Format (SAIF) for FPGA. Using these files we measured the dynamic (Vector Based) power consumption on the Synopsys PrimeTime and Vivado. Based on the measurement, the power consumption decreases quasi linearly with the entropy of the model. Again, this trend is due to the fact that FantastIC4 supports (1) the efficient processing of compressed representations of the weight parameters, (2) efficient computation of 4bit non-zero values, and (3) efficient loading of repeated values from the FIFOs; all being properties that become more and more predominant as the entropy of the models parameters decreases.

\section{Conclusion}
\label{conclusion}

In this paper, we proposed a software-hardware optimization paradigm for maximally increasing the area and power efficiency of MLP models with state-of-the-art predictive performance. Firstly, we introduce a novel entropy-constrained training method for making the models highly compressible in size, which, in combination with FantastIC4s supports for the efficient on-chip execution of multiple compact representations, boosts the data movement efficiency of the parameters by up to $29\times$ (on average 10.55$\times$ across different models) as compared to the original models, and by 2$\times$ as compared to previous compression approaches. In addition, our particular training algorithm renders the models to be robust to 4bit quantization while inducing sparsity, properties that FantastIC4 exploits in order to increase further the power efficiency by 2.7$\times$ and area efficiency by 2.6$\times$. Finally, it implements an activation stationary data movement paradigm, as such increasing the on-chip data movement efficiency of the activation values by 15$\times$. FantastIC4 was implemented on a Virtual Ultrascale FPGA XCVU440 device. The experimental results show that we achieve an overall throughput of 2.45 TOPS with a total power consumption of 3.6W. We achieved the lowest resource utilization for an multilayer perceptrons (MLPs) inference by consuming 67.24\% of  LUTs, 27.86\% of LUTRAMs, 3.88\% of FFs, 3.13\% of BRAMs and 0.27\% of DSPs. This is the first accelerator to achieve a very high throughput with a low resource utilization and a low power consumption up to date on an FPGA. We further benchmarked our FantastIC4 on a 22nm process, the ASIC version achieved a total power efficiency of 20.17 TOPS/W and a latency of 1.31$\mu$s per layer inference of the Google Speech Command (GSC) dataset. When compared to the other state of the art GSC accelerators, FantastIC4 is better by 51$\times$ in terms of throughput and 145$\times$ in terms of area efficiency.

\section*{Acknowledgements}
This work was supported by the Bundesministerium für Bildung und Forschung through the BIFOLD - Berlin Institute for the Foundations of Learning and Data (ref. 01IS18025A and ref 01IS18037A). The authors would further like to thank Global foundries (GF) for providing 22nm FDSOI MPW support.

\bibliography{References}
\bibliographystyle{IEEEtran}


\end{document}